\documentclass[amsmath, amsfonts, showpacs, superscriptaddress,twocolumn,prx]{revtex4-1}
\usepackage{graphicx}
\usepackage{epsfig}
\usepackage{bm}
\usepackage{dcolumn}
\usepackage{amsmath}
\usepackage{amssymb}
\usepackage{color}

\def\be{\begin{equation}}
\def\ee{\end{equation}}
\def\vep{{\varepsilon}}

\begin{document}

\title{Topological supercurrents interaction and fluctuations \\ in the multiterminal Josephson effect}

\author{Hong-Yi~Xie}
\affiliation{Department of Physics, University of Wisconsin--Madison, Madison, Wisconsin 53706, USA}
\affiliation{Kavli Institute of Theoretical Sciences, University of Chinese Academy of Sciences, Beijing, 100190, China}
\author{Alex Levchenko}
\affiliation{Department of Physics, University of Wisconsin--Madison, Madison, Wisconsin 53706, USA}
\begin{abstract}
We study the Josephson effect in the multiterminal junction of topological superconductors. We use the symmetry-constrained scattering matrix approach to derive band dispersions of emergent sub-gap Andreev bound states in a multidimensional parameter space of superconducting phase differences. We find distinct topologically protected band crossings that serve as monopoles of finite Berry curvature. Particularly, in a four-terminal junction the admixture of $2\pi$ and $4\pi$ periodic levels leads to the appearance of finite-energy Majorana-Weyl nodes. This topological regime in the junction can be characterized by a quantized nonlocal conductance that measures the Chern number of the corresponding bands. In addition, we calculate current-phase relations, variance, and cross-correlations of topological supercurrents in multiterminal contacts and discuss the universality of these transport characteristics. At the technical level these results are obtained by integrating over the group of a circular ensemble that describes the scattering matrix of the junction. We briefly discuss our results in the context of observed fluctuations of the gate dependence of the critical current in topological planar Josephson junctions and comment on the possibility of parity measurements from the switching current distributions in multiterminal Majorana junctions. 
\end{abstract}

\date{March 21, 2019}

\maketitle

\section{Introduction} 

The universality of conductance fluctuations (UCF) is the hallmark of mesoscopic physics \cite{BLA-JETPLett85,BLA-DEK-JETPLett86,Lee-Stone-PRL85,Lee-Stone-Fukuyama-PRB87}. 
This phenomenon emerges from the quantum coherence of electron trajectories and is sensitive to changes in
external magnetic field or gate voltage. At temperatures below the Thouless energy, $T<E_{\mathrm{Th}}$, which is related to the inverse dwell time for an electron to diffuse across the sample $E_{\mathrm{Th}}=D/L^2$, the root-mean-square (rms) value of conductance fluctuations saturates to the universal value of the order conductance quantum $\sim e^2/h$ as long as the characteristic sample size $L$ is smaller than the dephasing length $L<L_\phi$. Interaction effects in normal metals barely change the magnitude and universality of conductance fluctuations, although they are crucially important in determining the temperature dependence of dephasing effects and, in particular, $L_\phi$ \cite{Aleiner-Blanter-PRB02}. The robustness of UCF can be rooted to the random matrix theory description of Wigner-Dyson statistics of electron energy levels in disordered conductors \cite{Dyson}. Indeed, in the Landauer picture of transport across a mesoscopic sample, conductance is given by $e^2/h$ times the number of single-particle levels within the energy strip of the width of Thouless energy. While the average number of such levels depends on the dimensionality, random matrix theory predicts that their fluctuation is universally of the order of one \cite{Altshuler-Shklovskii-JETP86,Mehta}. 

When superconductivity is induced at the boundary of the mesoscopic sample via the proximity effect, the universality of fluctuations remains intact \cite{Exp-UCF-SN-1,Exp-UCF-SN-2}. Indeed, the magnitude of sample-to-sample conductance fluctuations changes only by a numerical factor of the order of unity whose value depends on the underlying symmetry \cite{Brouwer-PhD,Beenakker-RMP}. Interestingly, the universality of fluctuations extends beyond conductance as it also manifests in the Josephson current of a superconducting-normal-superconducting (SNS) bridge. Indeed, extending the original ideas of Altshuler and Spivak \cite{BLA-BZS-JETP87}, who argued that random shifts of sub-gap energy levels with superconducting phase difference would alter the current, Beenakker showed \cite{Beenakker-PRL91} that in short junctions, $L\ll\xi$, where $\xi$ is the superconducting coherence length, the rms value of critical current fluctuations saturates to a universal bound $\sim e\Delta/h$ determined only by the superconducting energy gap $\Delta$ in the leads. Further a complete characterization of the supercurrent variance as a function of phase across the point contact Josephson junction was computed by Chalker and Mac\^edo \cite{Chalker}. In long junctions, $L\gg\xi$, supercurrent fluctuations cease to be universal and scale with $\sim e E_{\mathrm{Th}}/h$. However, a remarkable property of these fluctuations is that there is a regime where the entire critical current through the junction can be determined by the mesoscopic contribution when the average current is suppressed. 

In recent years the interest in Josephson physics has shifted towards junctions whose elements include topological materials \cite{Sacepe,Veldhorst,Bestwick,Mason,Finck-PRX,Kurter-PRB,Kurter-NC,Sochnikov,Stehno} or where topological properties are enabled by a specific design of the hybrid junction with otherwise conventional materials \cite{Frolov,Shtrikman,Marcus,Moler,Kouwenhoven}. These possibilities and advances motivate our work to investigate how universal mesoscopic effects manifest in topological Josephson junctions that, in particular,  host Majorana states (see the review in \cite{Beenakker-RMP-MF} and references therein). We carry out this analysis in the context of multiterminal devices that were brought into the spotlight of recent theoretical attention with the observation that they can emulate topological matter \cite{Riwar,Eriksson,Houzet,HYX1,HYX2,Qi,Nazarov,Deb,Meyer}, which triggered experimental efforts in realizing these systems in various proximitized circuits \cite{Giazotto,Heiblum,Finkelstein,Manucharyan,Pribiag}.          

The rest of the paper is organized as follows. In Sec. \ref{sec:scattering} we briefly review symmetry-constrained scattering matrix transport formalism in application to the Josephson effect in multiterminal circuits. In Sec. \ref{sec:junctions} we apply these methods to two-terminal junctions as a benchmark and then extend our analysis to three- and four-terminal devices, for which we compute the emergent band structure of sub-gap states, investigate their topology, and derive transport characteristics such as transconductance and supercurrent. In Sec. \ref{sec:fluctuations} we focus our attention on the statistical properties of topological supercurrents and obtain analytical results for variance that takes a universal form and also inherits the $4\pi$ periodicity of the Majorana Josephson effect.         


\section{Scattering matrix formalism}\label{sec:scattering}

Consider a Josephson junction (JJ) where $n$ superconducting (S) terminals are connected through the common normal (N) region, thus forming a multiterminal SNS contact. To keep the presentation simple, we assume that each superconducting lead is coupled by only a single conducting channel in the normal region and both time-reversal and chiral symmetries are broken (unconventional classes D and C \cite{Beenakker-RMP-MF}). Formation of the sub-gap bound states in the JJs is the result of coherent Andreev reflections that describe electron-to-hole conversion at the superconductor-normal interface. In $n$-terminal junctions an elastic scattering event at energy $\vep$ is characterized by a scattering matrix $\hat{S}(\vep) \in \mathrm{U}(2n)$, where ``$2$'' denotes the particle-hole degrees of freedom. In what follows we assume that all leads have the same superconducting gap $\Delta$ and normalize all energies in units of $\Delta$. The particle-hole (PH) symmetry is represented by 
\be  \label{PHS}
\hat{S}(\vep) = \mathcal{P} \hat{S}(-\vep) \mathcal{P}^{-1},
\ee 
where the antiunitary PH transform $\mathcal{P}$ falls into two categories, $\mathcal{P}^2 = \pm 1$. For example, for $s$-wave paring, $\mathcal{P} = \hat{\tau}_1 \mathcal{K}$ ($\mathcal{P}^2 =  +1$) in the spin-nondegenerate case, and $\mathcal{P} = i \hat{\tau}_2 \mathcal{K}$ ($\mathcal{P}^2=-1$) in the spin-degenerate case, where $\hat{\tau}_{1,2,3}$ are the Pauli matrices acting in particle-hole space and $\mathcal{K}$ denotes the complex conjugation. The Andreev bound state energies are determined by the determinant equation \cite{Beenakker-PRL91}
\be \label{det-0}
\mathrm{Det}[\mathbb{I}_{2n}-\hat{R}_A (\vep,\hat{\theta}) \hat{S}_N(\vep)]= 0.
\ee
Here $\hat{S}_N(\vep)$ is the scattering matrix of the normal region, and $\hat{R}_A (\vep,\hat{\theta})$ is the scattering matrix describing Andreev reflections, where $\theta_\alpha \in \{ \theta_0, \theta_1, \cdots, \theta_{n-1}\}$ is the diagonal matrix of superconducting phases. We set $\theta_0 = 0$ owing to global gauge invariance. Due to the PH symmetry Eq. \eqref{PHS} these scattering matrices take the block-diagonal forms
\begin{align}
&\hat{S}_N(\vep) = \begin{bmatrix} \hat{s}(\vep) & 0 \\ 0 & \hat{s}^\ast(-\vep) \end{bmatrix}, \nonumber \\ 
&\hat{R}_A(\vep,\hat{\theta}) = e^{-i\arccos{\vep}} \begin{bmatrix} 0 & e^{i \hat{\theta}} \\  -\mathcal{P}^2 e^{-i \hat{\theta}} & 0 \end{bmatrix},
\end{align}
where $\hat{s}(\vep) \in \mathrm{U}(n)$. The determinant in Eq.~\eqref{det-0} simplifies further to a degree-$n$ characteristic polynomial of $\gamma(\vep) \equiv e^{-2i\arccos{\vep}}$,
\be \label{det-2}
P_{n}(\gamma;\hat{\theta},\vep) \equiv \mathrm{Det}\left[ \mathbb{I}_{n} + \mathcal{P}^2 \gamma(\vep) e^{i \hat{\theta}} s^\ast(-\vep) e^{-i \hat{\theta}} s(\vep) \right]. 
\ee
which is (anti)palindromic $\mathcal{P}^{2n} \gamma^n P_n(\gamma^{-1}) = P_n(\gamma)$. Importantly, from Eq.~\eqref{det-2} we observe that, for a fixed normal-region scattering matrix $\hat{s}$, the Andreev bands of $\mathcal{P}^2= \pm 1$ symmetry classes are dual via the relation 
\be \vep_{\mathcal{P}^2=+1}^2(\hat{\theta})+\vep_{\mathcal{P}^2=-1}^2(\hat{\theta})=1. 
\ee
Previously, we extensively discussed the $\mathcal{P}^2=-1$ scenario in application to three- and four-terminal junctions \cite{HYX1,HYX2}. In this work we primarily focus on the Andreev spectrum of $\mathcal{P}^2=+1$ junctions that can support zero-energy Majorana modes.  In what follows, we also assume energy-independent scattering matrices $\hat{s}$ that correspond to, for example, weak links where the length of the junction is small compared to the superconducting coherence length, $L\ll\xi$, so that retardation effects of traveling quasiparticles can be neglected. We note that the existence of Majorana zero modes does not depend on this assumption.

To study the energy spectra of emergent states in junctions with $\mathcal{P}^2=+1$ terminals, we introduce the scattering matrix at $\vep =0$,
\be  \label{e0S}
\hat{S}_0 \equiv \hat{R}_A (0,\hat{\theta}) \hat{S}_N= i\begin{bmatrix} 0 & - e^{i \hat{\theta}} \hat{s}^\ast \\  e^{-i \hat{\theta}} \hat{s}  & 0  \end{bmatrix},
\ee 
which belongs to the circular real ensemble since $\mathrm{Det} \hat{S}_0 = (-1)^n$. Via Eq.~\eqref{det-0} the zero-energy Majorana modes are determined by the determinant equation of an antisymmetric matrix $\hat{m}(\theta)$,
\be  \label{Det-e0}
\mathrm{Det}[\hat{m}(\hat{\theta})]=0, \quad \hat{m}(\hat{\theta})  = e^{-i\hat{\theta}/2} \hat{s} e^{i\hat{\theta}/2}-e^{i\hat{\theta}/2} \hat{s}^{T} e^{-i\hat{\theta}/2}.
\ee
From here we draw important properties. (i) For $n \in \mathrm{odd}$, Eq.~\eqref{Det-e0} is generally satisfied for any scattering matrices $\hat{s}$ and phases $\hat{\theta}$. This implies that Andreev-Majorana zero modes are present at any phases and robust to elastic scattering and superconducting order parameter nonuniformity. These nondispersive flat bands do not contribute to Josephson currents. (ii) For $n \in \mathrm{even}$, the Andreev bands cross at zero energy at phases determined by the Pfaffian equation 
\be \label{pf}
\mathrm{Pf}_{n \in \mathrm{even}}[\hat{m}(\hat{\theta})]=0.
\ee
Based on our study of two- and four-terminal junctions, we conjecture that there always exists a pair of Majorana zero-modes modes on an $(n-2)$-dimensional hypersurface in the $\boldsymbol{\theta}=(\theta_1, \cdots, \theta_{n-1})$ space described by Eq.~\eqref{pf}. Next we reveal the energy spectrum of the junction for several concrete forms of the scattering matrix.  

\begin{figure}
\includegraphics[width=0.235\textwidth]{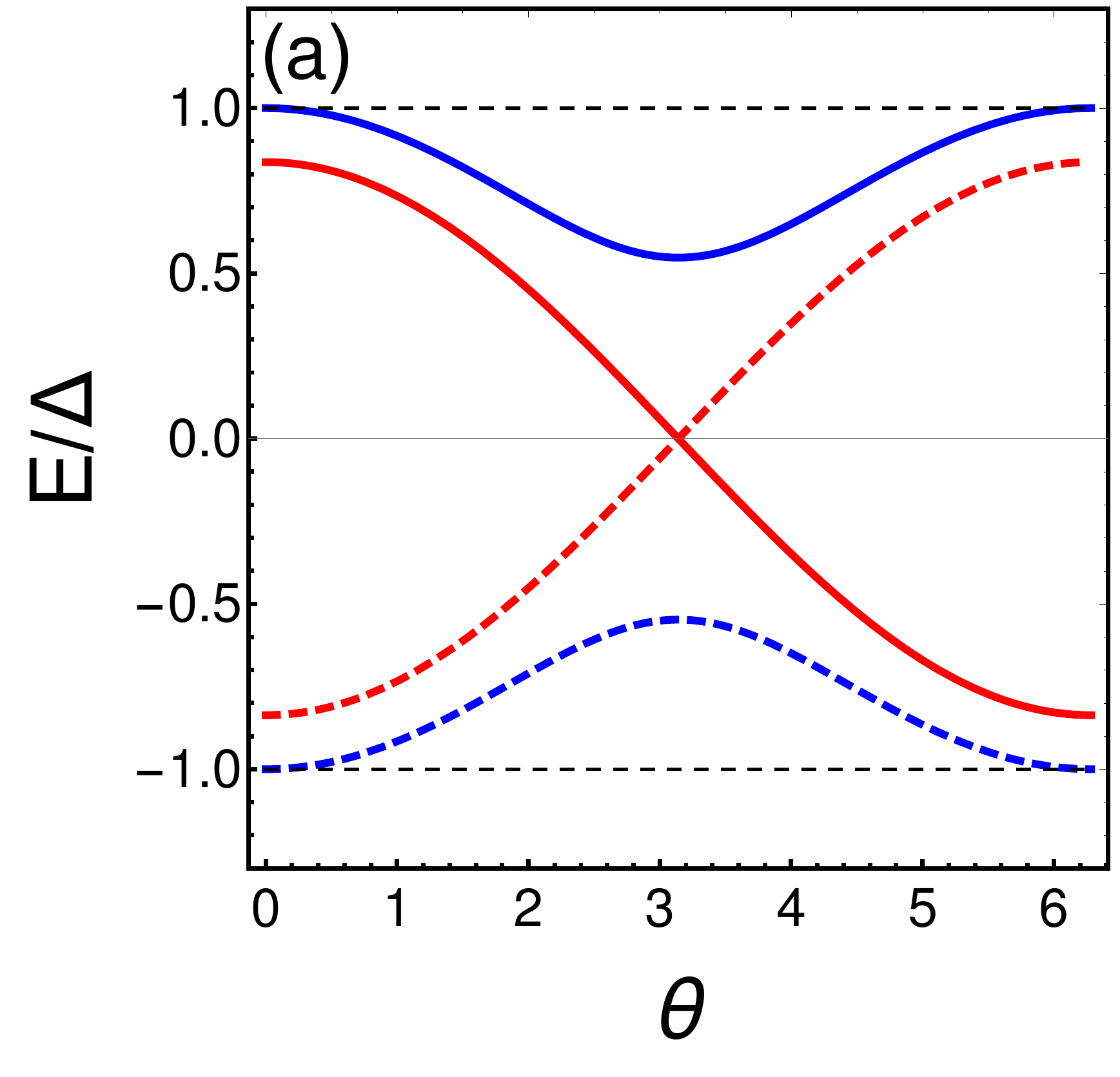}
\includegraphics[width=0.235\textwidth]{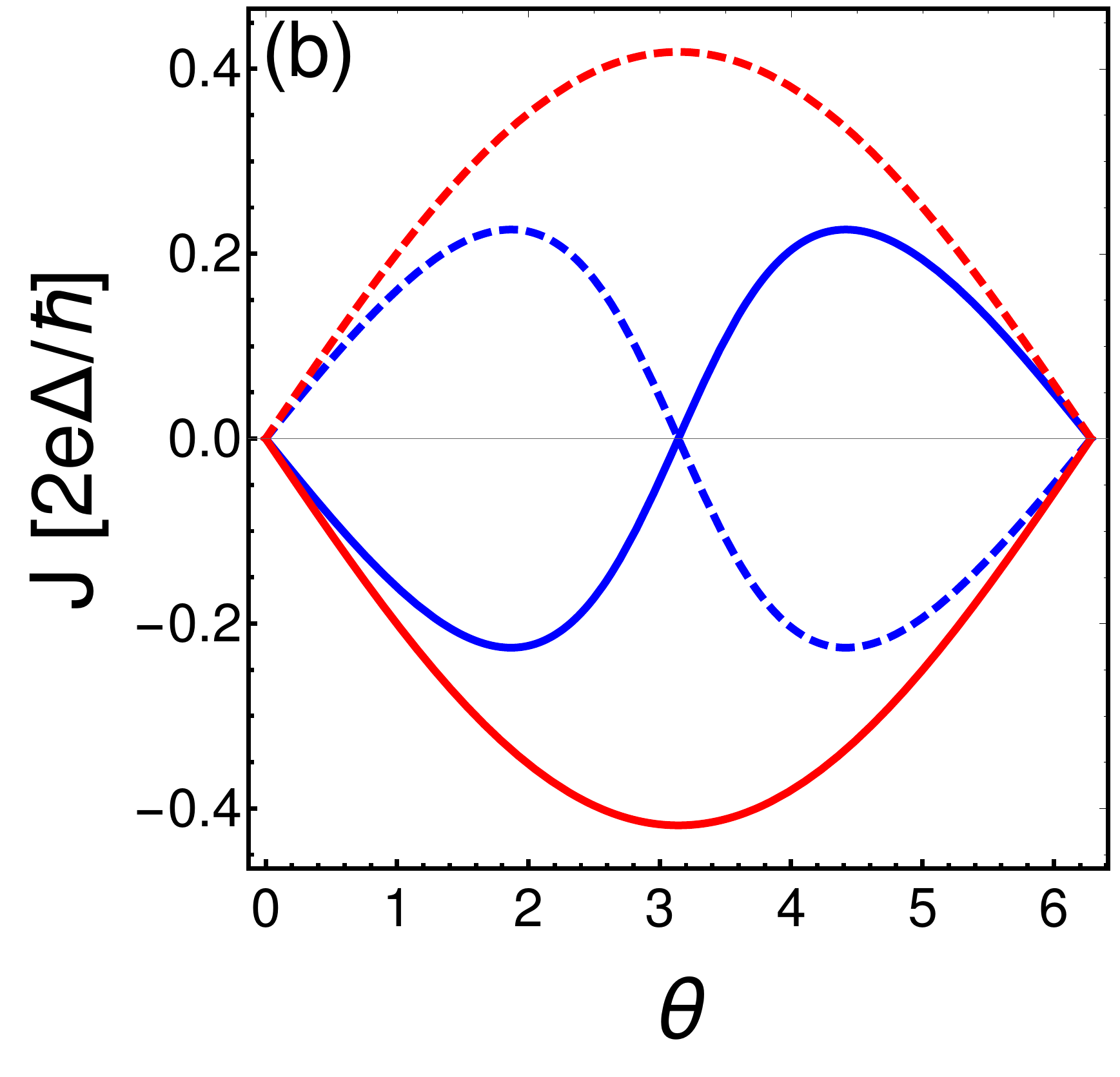}
\caption{[Color online] (a) Energy spectrum [Eq.~\eqref{2T-spec}] and (b) Josephson current [Eq.~\eqref{2T-J}] for $\mathcal{P}^2 = +1$ two-terminal topological Josephson junctions. We take $T=0.7$ and $\phi=\pi$. The red (blue) curves are the results for topologically nontrivial leads $\mathcal{P}^2 = +1$ (topologically trivial leads $\mathcal{P}^2 = -1$). In panel (a) the black dashed lines indicate the continuum edge $\vep=E/\Delta= 1$.} 
\label{fig-2T}
\end{figure}


\section{Multiterminal Josephson effect}\label{sec:junctions}

\subsection{Two-terminal junctions}

We first study two-terminal junctions as a benchmark. We parametrize the $2\times 2$ unitary matrix $\hat{s}$ by four independent parameters,
\be \label{s2}
s = 
\begin{bmatrix} 
\sqrt{1-T} e^{i \varphi_{00}}  &  \sqrt{T}   e^{i \varphi_{01}} \\ 
\sqrt{T}   e^{i \varphi_{10}}  &  -\sqrt{1-T} e^{i (\varphi_{01}+\varphi_{10}-\varphi_{00})} 
\end{bmatrix}, 
\ee
where $T \in [0,1]$, representing the normal-region transmission and scattering phases $\varphi_{00,01,10} \in [0, 2 \pi]$. The sub-gap spectrum of excitations is determined by the $n=2$ characteristic polynomial \eqref{det-2} via the equation $P_2(\gamma) = \gamma^2 + 2 B_2 \gamma + 1=0$, where the $B_2$-function takes the form $B_2(\theta)  = 1 - 2 T \sin^2(\vartheta/2)$, with $\vartheta \equiv \theta - \phi+\pi$ and $\phi \equiv \varphi_{10}-\varphi_{01}$. The two branches of dispersive solutions are given by 
\be \label{2T-spec}
\vep(\vartheta) = 
\pm\begin{cases}
\sqrt{T}  \cos(\vartheta/2), & \quad \mathcal{P}^2 = +1, \\
\sqrt{1-T \sin^2(\vartheta/2)}, & \quad \mathcal{P}^2 = -1,
\end{cases}
\ee
where for comparison we recall the results for the conventional $\mathcal{P}^2=-1$ junctions. For finite transmission $T \neq 0$, the $n=2$ Pfaffian equation \eqref{pf} reduces to $\mathrm{Pf}_2(\theta)= m_{01} \sim \cos(\vartheta/2)=0$, so that a Majorana crossing occurs at $\vartheta=(2k+1)\pi$, with $k \in \mathbb{Z}$. The zero-temperature Josephson current $J(\theta) \equiv (2 e \Delta/\hbar) \partial_{\theta} \vep$ takes the form 
\be \label{2T-J}
J(\vartheta) = \pm 
\frac{e \Delta}{\hbar}  \times \begin{cases} 
\sqrt{T} \sin(\vartheta/2), & \quad \mathcal{P}^2 = +1, \\
T \sin{\vartheta}/4\vep(\vartheta), & \quad  \mathcal{P}^2 = -1.
\end{cases} 
\ee
The typical energy dispersion and supercurrent-phase relation are shown in Fig.~\ref{fig-2T}. We note that for $\phi=\pi$ doubly degenerate Majorana states emerge at $\theta = \pi$ and the energy and supercurrent exhibit $4 \pi$ periodicity in $\theta$. In addition, the bound states are detached from the continuum with a minimal gap $1- \sqrt{T}$ at $\theta = 0$ and $2\pi$. Equations \eqref{2T-spec} and \eqref{2T-J} are consistent with the prior results (e.g., Ref. \cite{Kwon}).  

\begin{figure}
\includegraphics[width=0.23\textwidth]{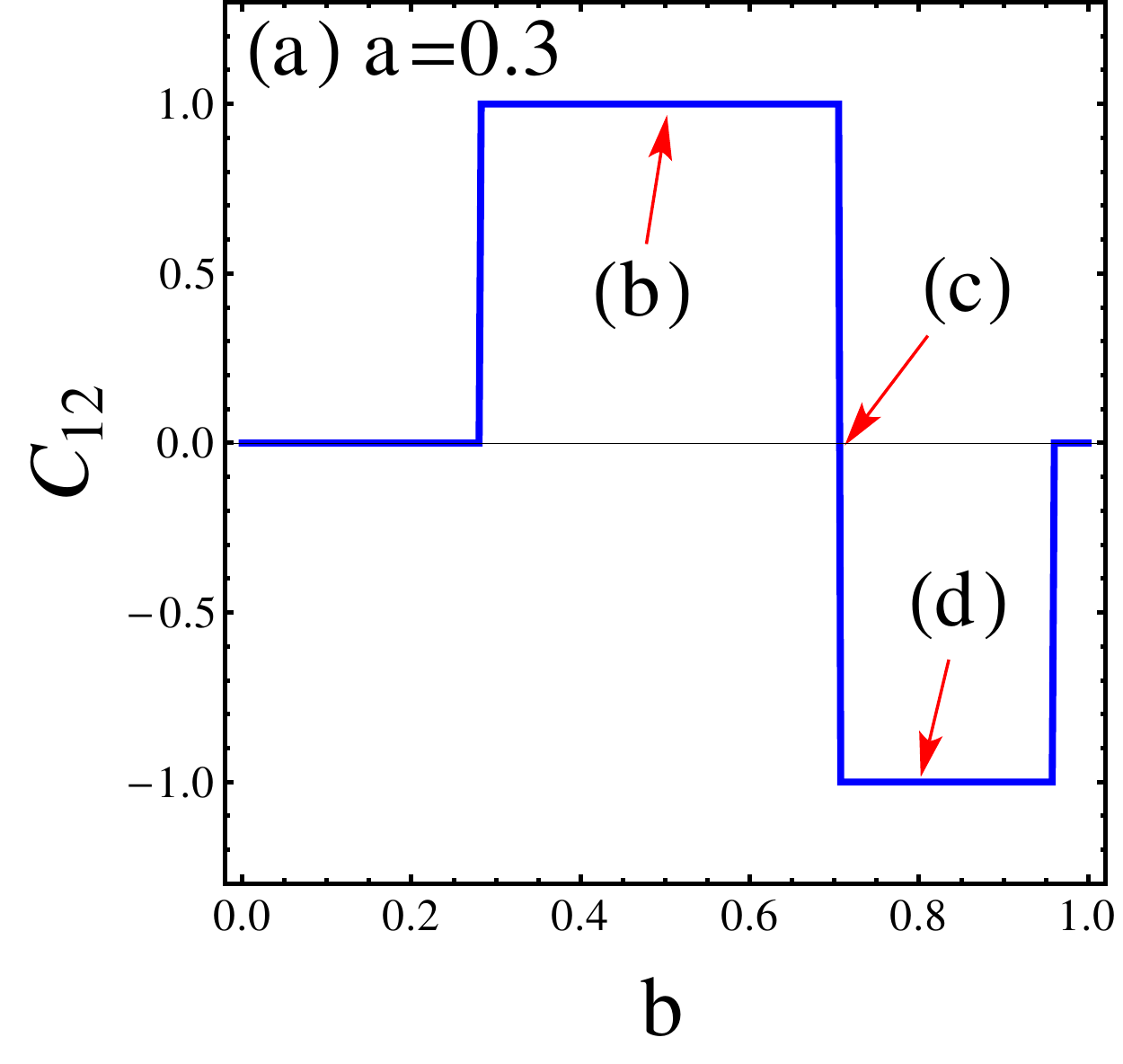} 
\includegraphics[width=0.23\textwidth]{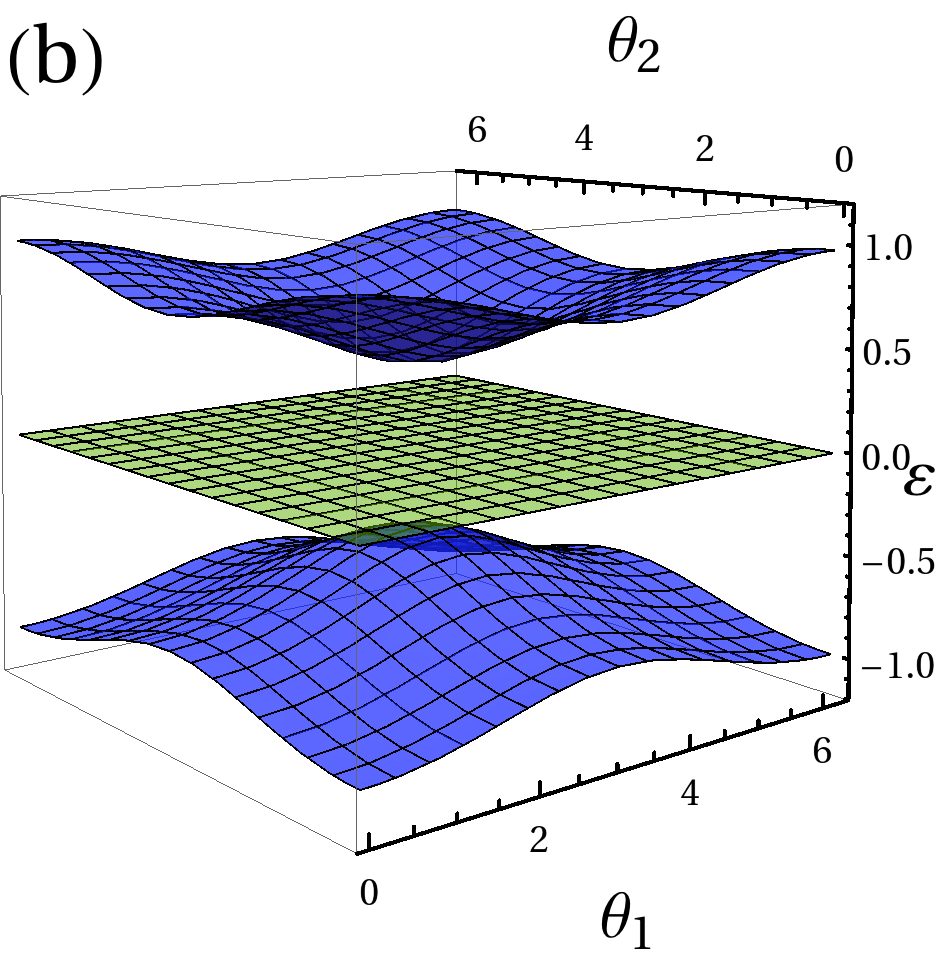} \\
\includegraphics[width=0.23\textwidth]{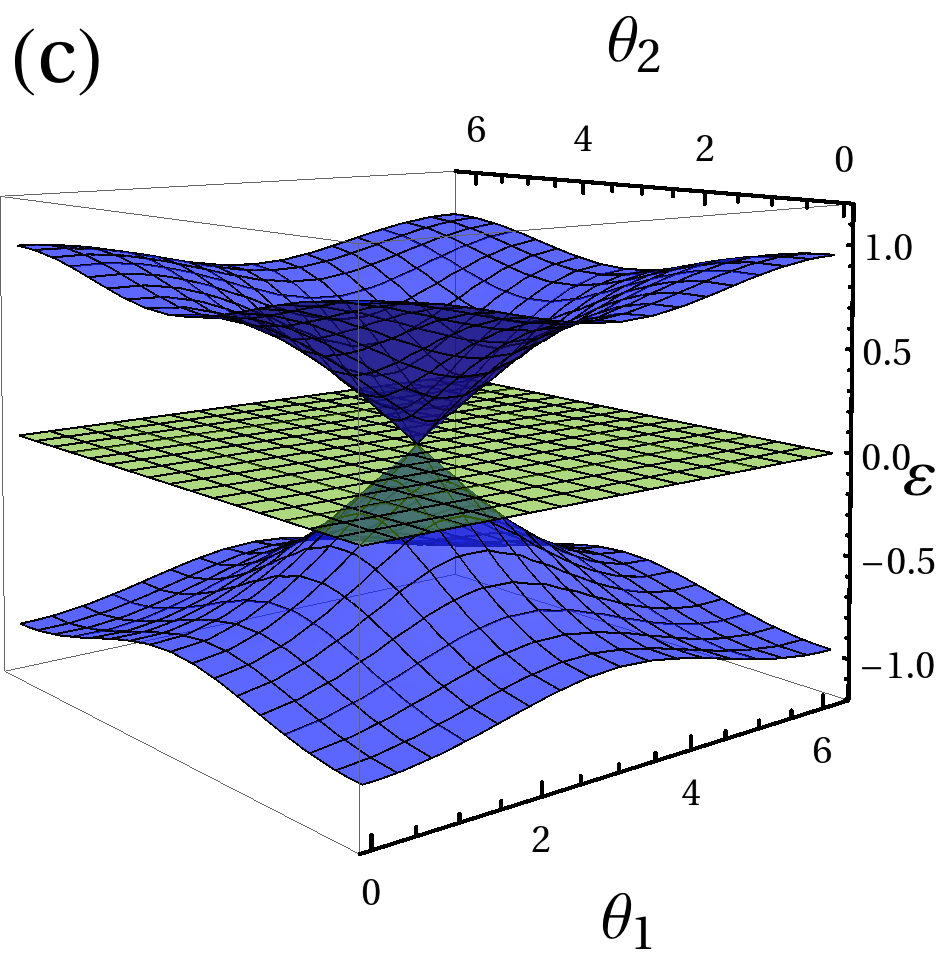} 
\includegraphics[width=0.23\textwidth]{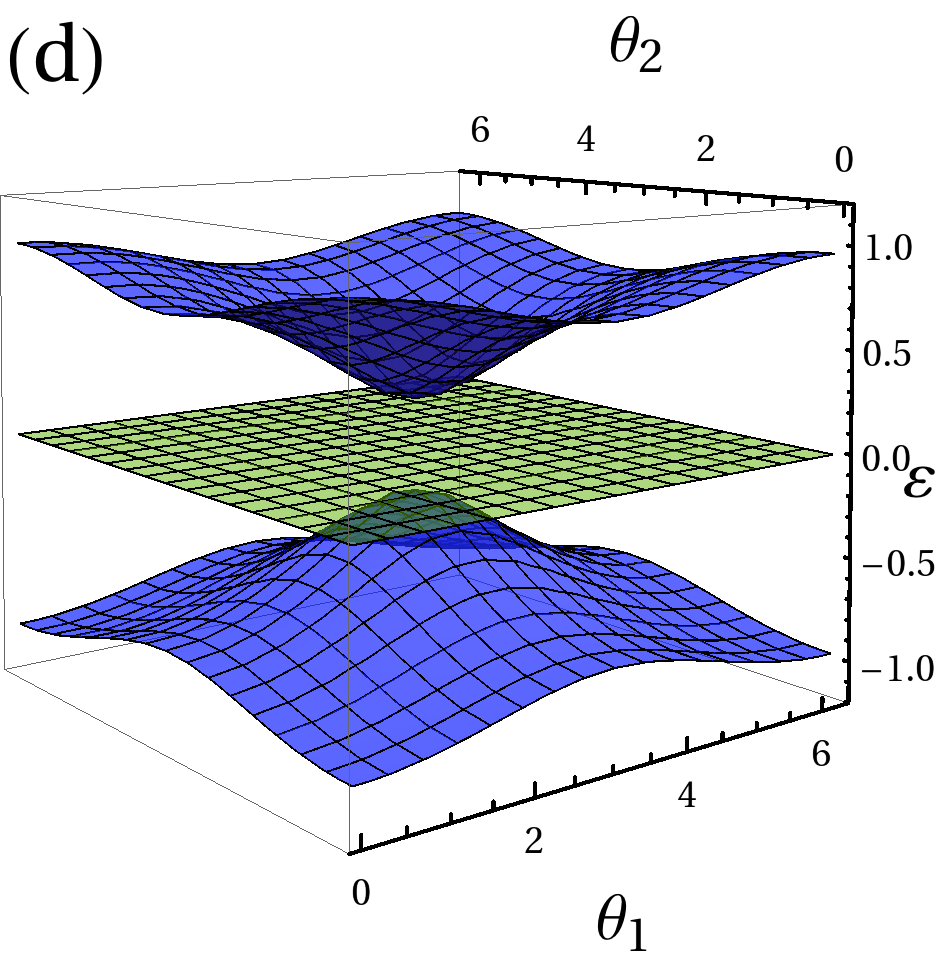} \\
\includegraphics[width=0.22\textwidth]{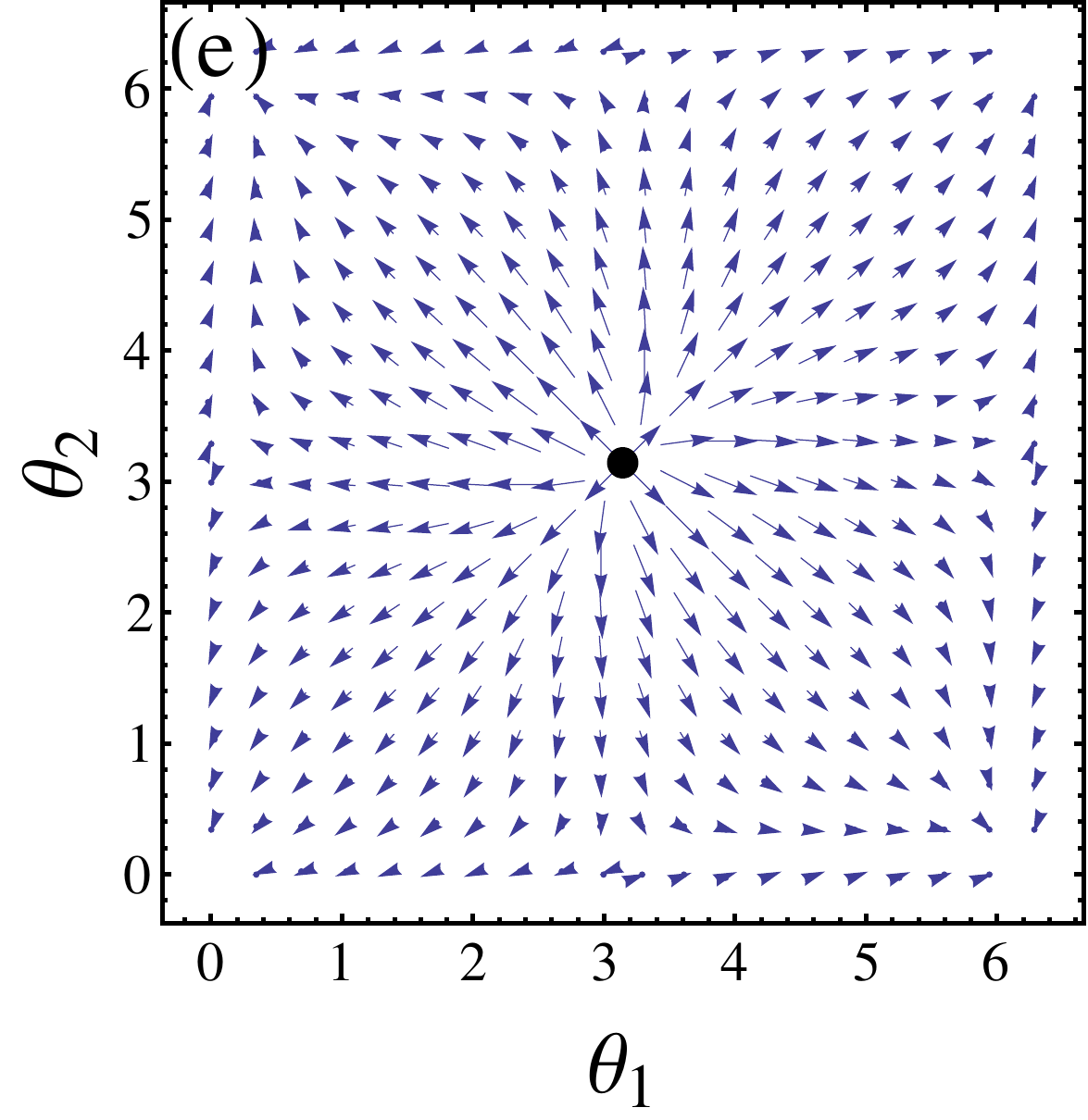} 
\includegraphics[width=0.23\textwidth]{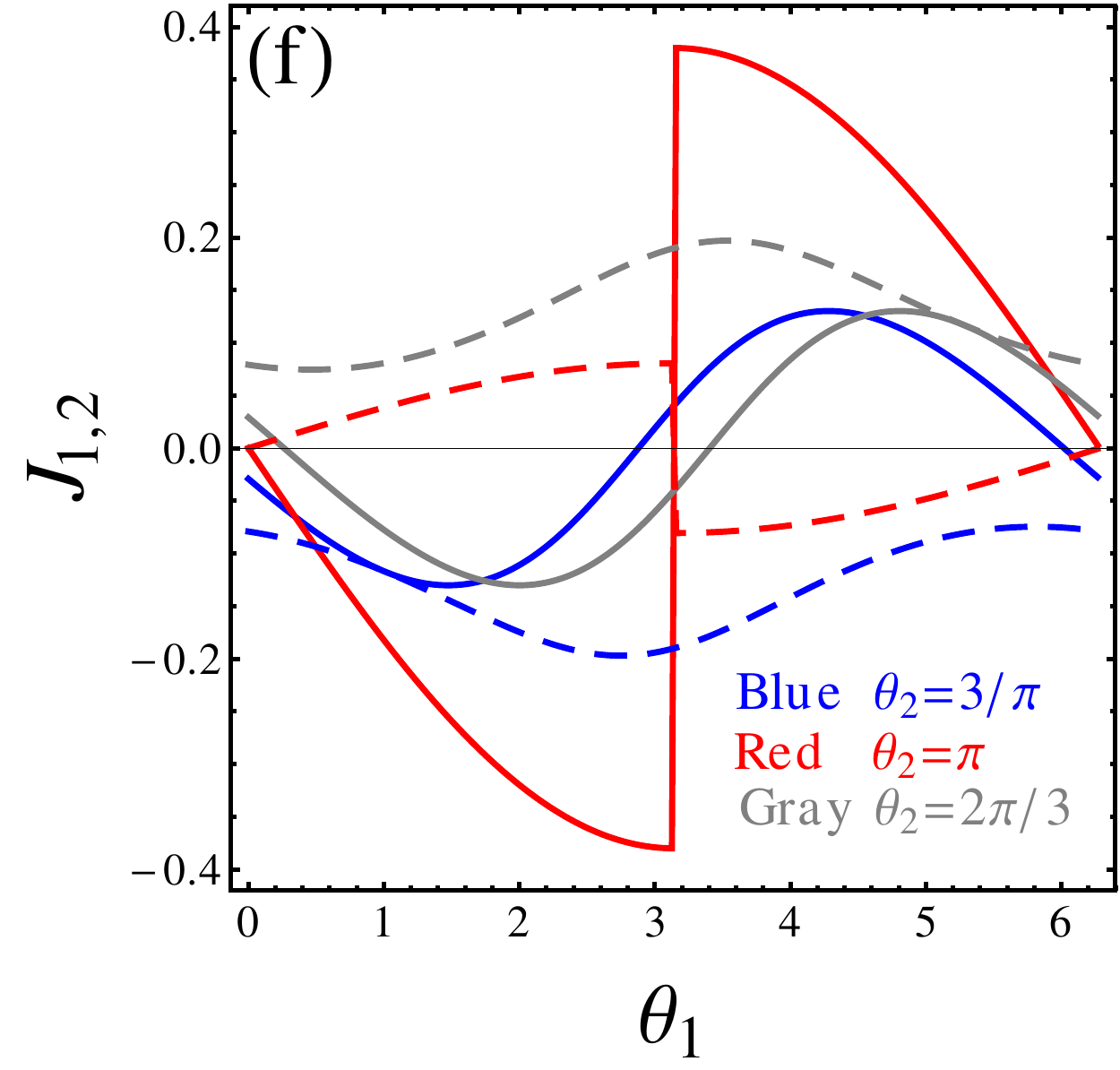}  
\caption{[Color online] Energy spectrum and Josephson current for $\mathcal{P}^2=1$ three-terminal junctions. 
We take $c=\sqrt{1-b^2}$, $a=0.3$, and $\varphi=\phi_1=\phi_2=\pi$.
(a) Chern number as a function of $b$. (b)-(d) Andreev spectra at $b=0.5$, $b_0=1/\sqrt{2}$, and $0.8$. 
(e) and (f) Josephson currents $J_{1,2}$ as functions of $\theta_{1,2}$ at $b=b_0$. Panel (e) shows a hedgehog-like pattern of the  current flow about the Weyl node. Panel (f) shows $J_{1,2}$ as a function of $\theta_1$ for various values of $\theta_2$.} 
\label{fig-3T}
\end{figure}

\subsection{Three-terminal junctions}

For $n=3$ the spectrum of localized states is determined by the palindromic polynomial $P_3(\gamma) =(\gamma+1)(\gamma^2 + 2 B_3 \gamma + 1)=0$ and composed of three bands, 
\be  \label{3T-spec}
\vep_\pm (\boldsymbol{\theta}) = \pm \sqrt{\frac{1-B_3(\boldsymbol{\theta})}{2}}, \quad \vep_0(\boldsymbol{\theta})=0. 
\ee
Adopting the same parametrization of the scattering matrix as in Ref. \cite{HYX1}, the $B_3$-function can be found in the closed analytical form
\begin{align} 
&B_3= \frac{1}{2}\left[ 2 a^2 - (1+a^2)(b^2+c^2 - 2b^2c^2)\right. \nonumber\\ 
&\left. - 4 abc \sqrt{(1-b^2)(1-c^2)} \cos\varphi \right] \nonumber\\ 
&+ bc(1-a^2) \cos\vartheta_1 + (1-a^2) \sqrt{(1-b^2)(1-c^2)} \cos\vartheta_2  \nonumber\\
&+\left[bc (1+a^2) \sqrt{(1-b^2)(1-c^2)} + a (b^2+c^2 - 2b^2c^2) \cos\varphi \right]\nonumber \\ 
&\times \cos(\vartheta_1-\vartheta_2)+a (b^2-c^2) \sin\varphi \sin(\vartheta_1-\vartheta_2). \label{B-trsb} 
\end{align}
Consequently, there are only six independent parameters of the scattering matrix $\{ a, b, c, \varphi, \phi_{1,2} \}$ that enter the spectrum of Andreev bound states (ABS). Furthermore, scattering phases $\phi_{1,2}$ only shift the phases of the leads  $\vartheta_{1,2}=\theta_{1,2}-\phi_{1,2}$.

Depending on the choice of scattering matrix parameters, we find rich behavior of the energy bands. For the special case $c=\sqrt{1-b^2}$ and $\phi=\pi$ the spectrum exhibits nontrivial topology, as shown in Fig. \ref{fig-3T}. Zero-energy Weyl points appear at $\vartheta_{1,2}=0$ for $b=b_0=1/\sqrt{2}$ [Fig.~\ref{fig-3T}(b)-(d)]. As shown in Fig.~\ref{fig-3T}(a), the Chern number of the corresponding band structure exhibits a sign jump $C_{12} = \mathrm{sgn}(b_0-b)$ for $b \to b_0$. We also note that the other topological phase transitions for $b \approx 0.28$ and $0.96$ are related to the gap closing/reopening at the Andreev band edge $\vep=1$. Figure \ref{fig-3T}(e) displays Josephson currents $J_{1,2}$ in two terminals when the system is tuned to the nodal gapless states. In Fig.~\ref{fig-3T}(f) the series of one-dimensional cuts in either the $\theta_1$ phase or $\theta_2$ shows how Josephson currents change as one tunes phases to the vicinity of nodal points. We observe that moving across the node currents exhibit discontinuous jumps. Note that the Majorana flat band $\vep_0=0$ does not contribute to the Josephson current. 

\begin{figure}
\includegraphics[width=0.23\textwidth]{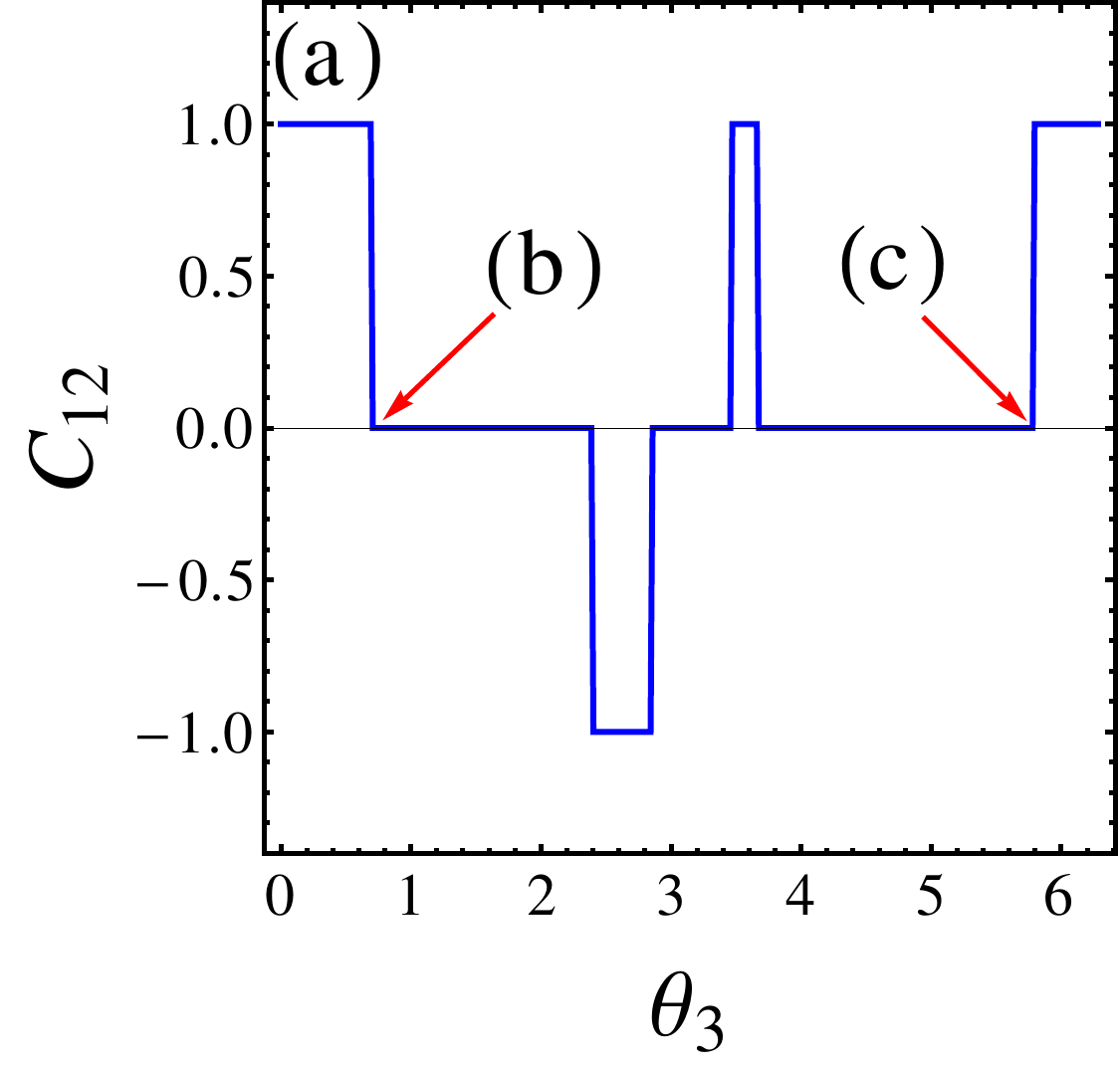} \,
\includegraphics[width=0.23\textwidth]{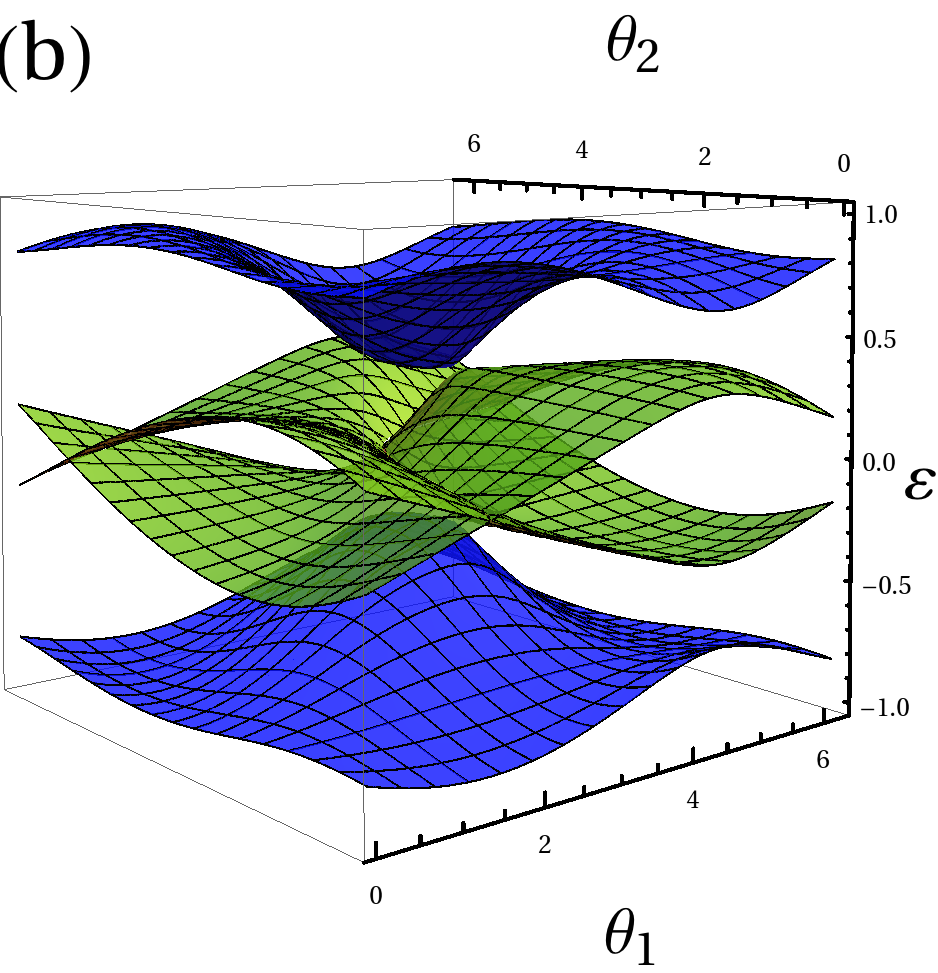} \\
\includegraphics[width=0.23\textwidth]{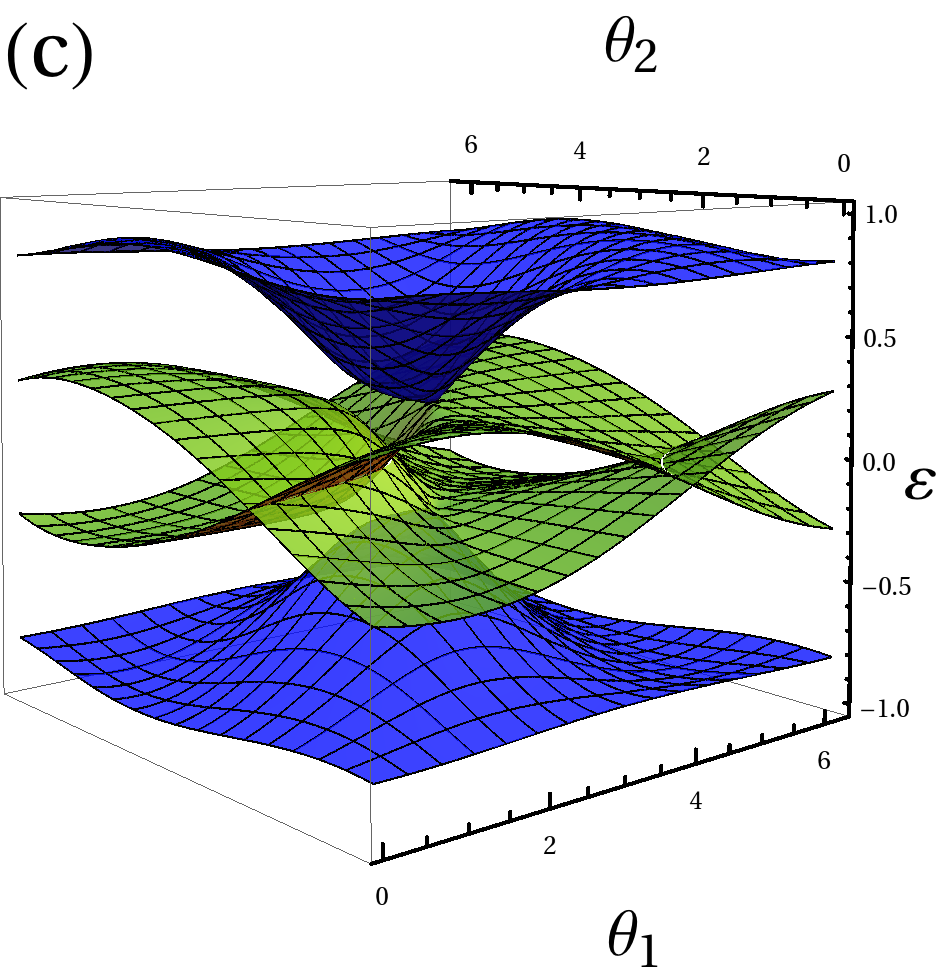} 
\includegraphics[width=0.23\textwidth]{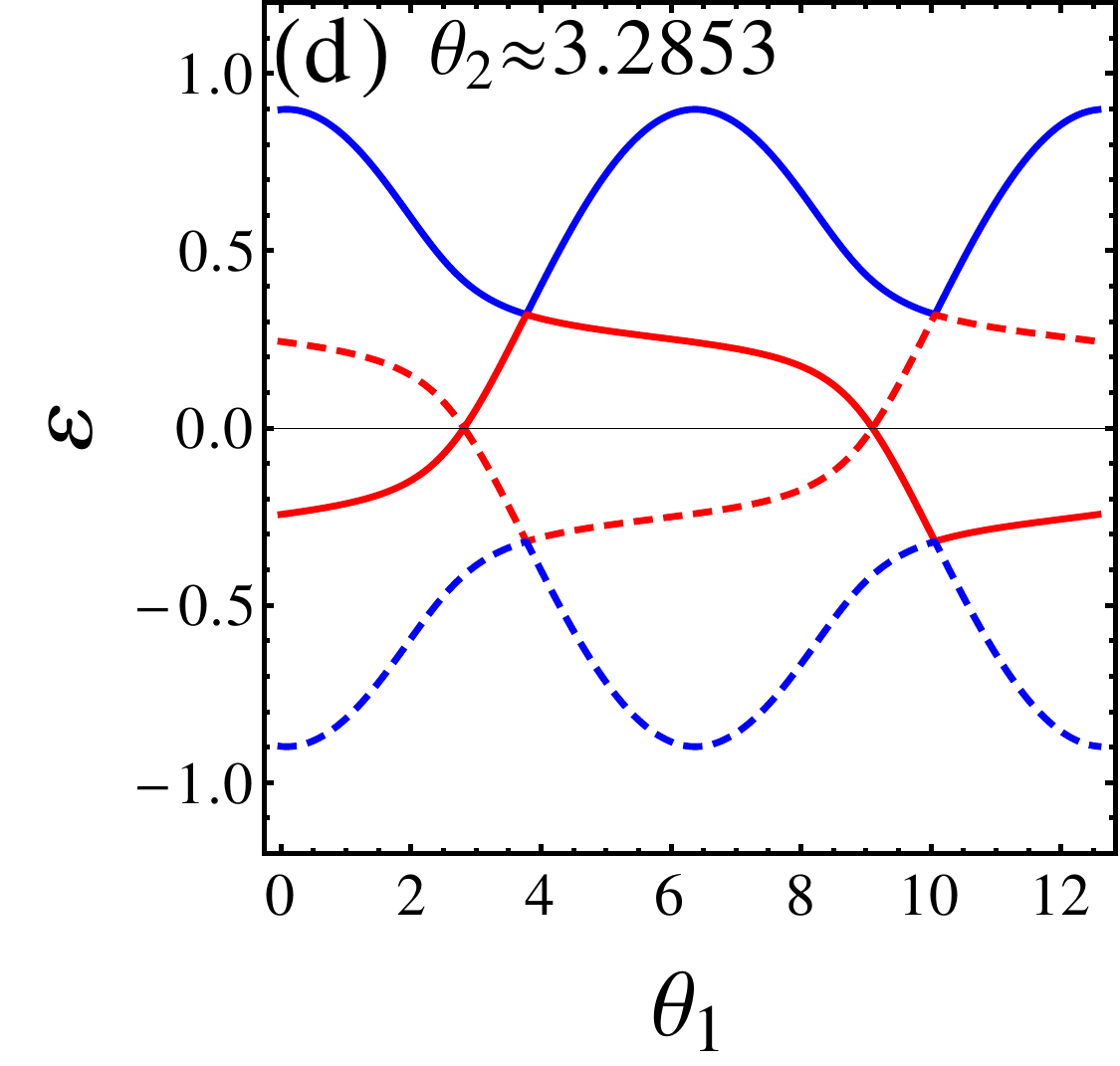} \\
\includegraphics[width=0.23\textwidth]{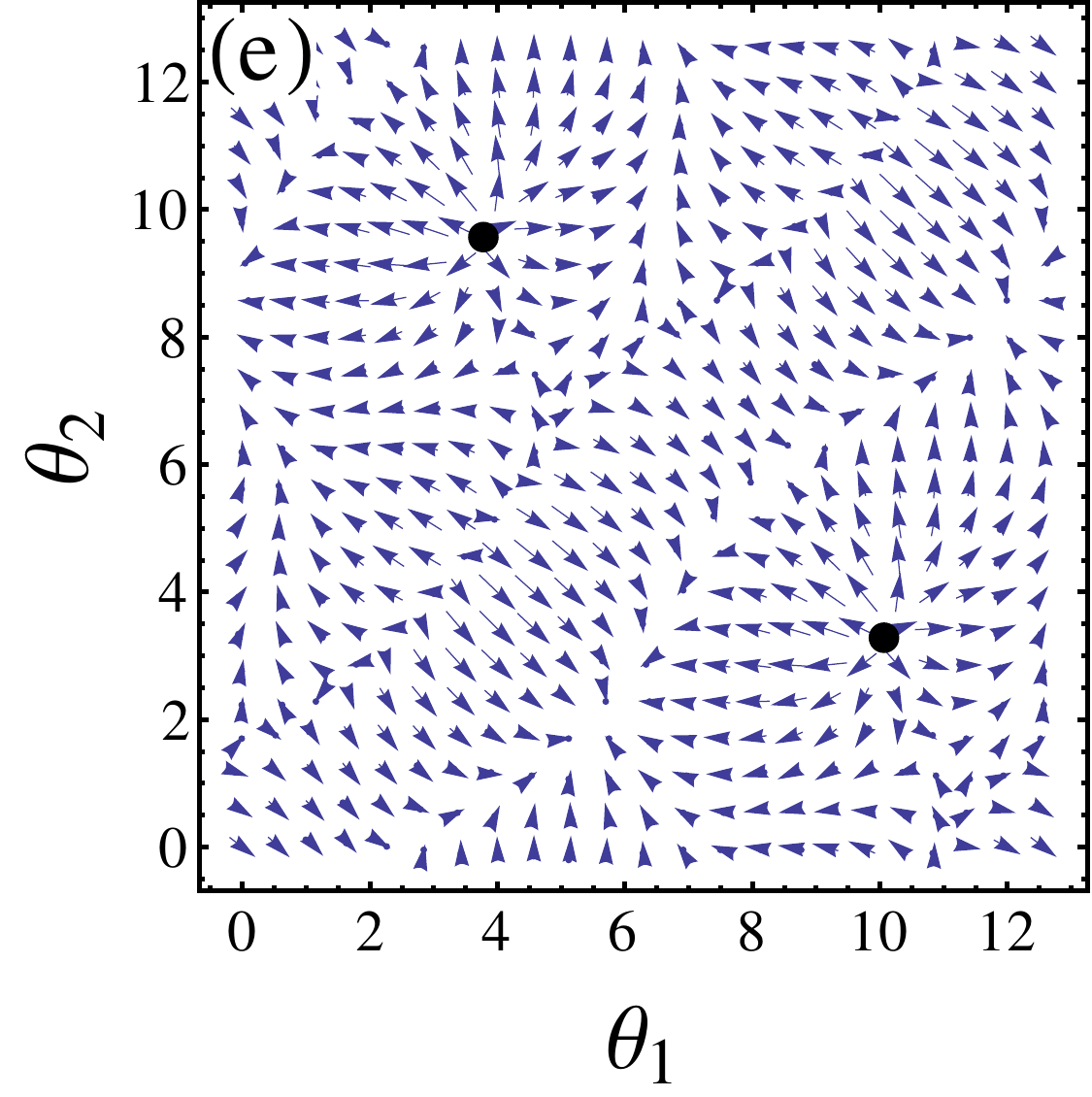}
\includegraphics[width=0.23\textwidth]{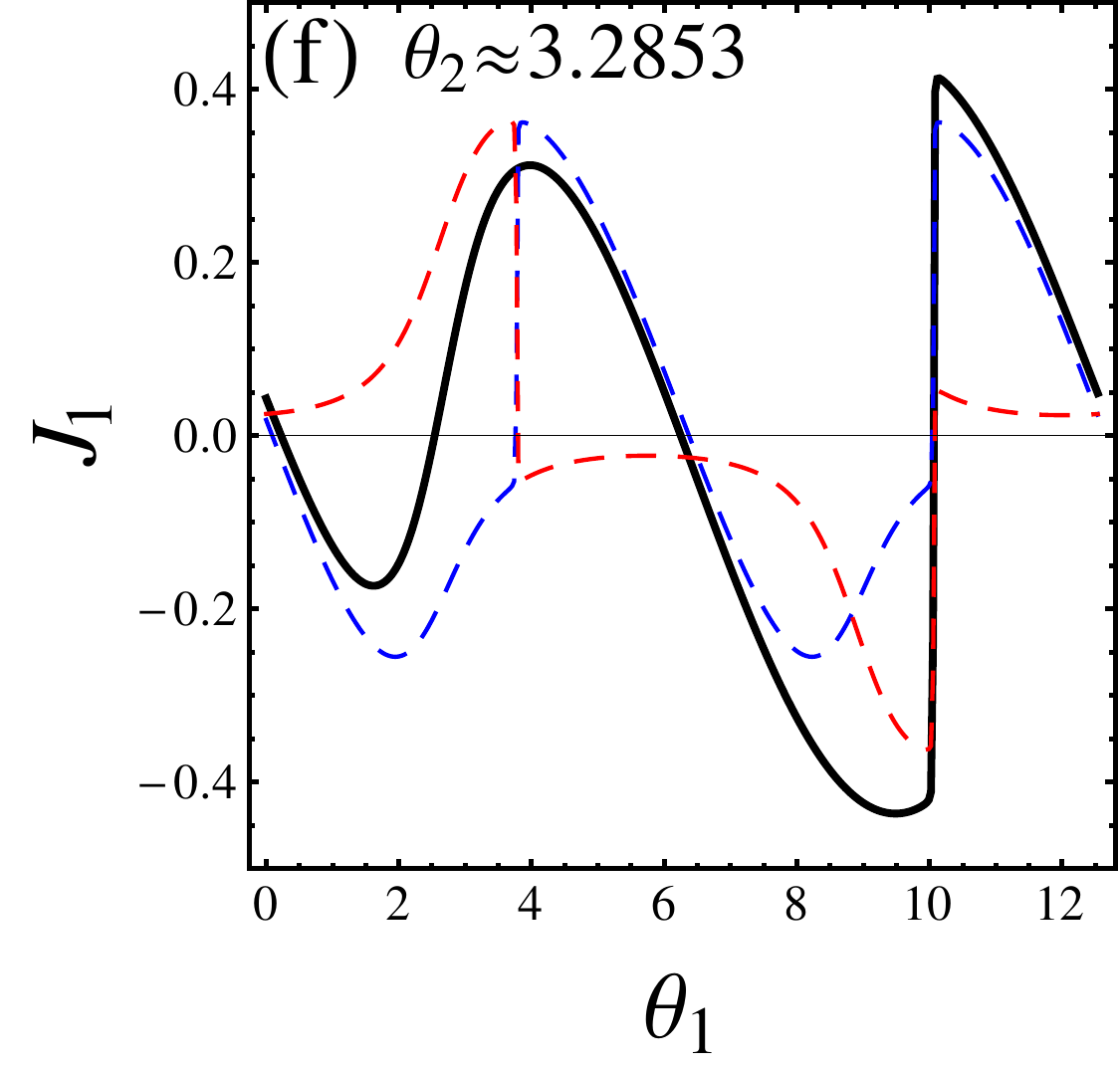}
\caption{[Color online] Energy spectrum for $\mathcal{P}^2=1$ time-reversal-broken four-terminal junctions. The scattering matrix parameters are defined as in Ref.~\cite{HYX1}. We take $a=1/4$, $b=1/\sqrt{3}$, $c=1/5$, $d=1/2$, $f=1/3$, $h=4/5$,  
$\phi_{1} = \phi_2=\pi$, $\phi_3=0$, $\phi_4=\phi_5=-\pi/3$, and $\phi_6=\pi/6$. 
(a) Chern number as a function of $\theta_3$. At $\theta_3^\ast \approx 0.6990$ and $5.7883$, finite-energy Weyl nodes form at $(\theta_1^\ast,\theta_2^\ast) \approx (3.7785,3.2853)$ and $(3.2046,3.0051)$ and the spectra are shown in panel (b) and (c), respectively. 
(d) Traces of the spectrum in panel (b) at $\theta_2 = \theta_2^\ast$. Upper (blue) and lower (red) Andreev bands exhibit $2\pi$ and $4 \pi$ periodicity, respectively. 
Panel (e) shows the pattern of Josephson currents $J_{1,2}(\theta_1,\theta_2)$ corresponding to the spectrum (b). The hedgehog-like singularities are present at the Weyl nodes $(\theta_1,\theta_2) \approx (\theta_1^\ast + 2 \pi,\theta_2^\ast)$ and $(\theta_1^\ast,\theta_2^\ast+2 \pi)$. (f) Trace of $J_{1}(\theta_1)$ at $\theta_2=\theta_2^\ast$. The dashed lines denote the contributions of upper (blue) and lower (red) bands.} 
\label{fig-4T}
\end{figure}

\subsection{Four-terminal junctions}

The energy spectrum of four-terminal junctions can host Majorana zero modes and Weyl nodes simultaneously. The four Andreev bands determined by the palindromic equation $P_4 (\gamma) = \gamma^4 + A_4 \gamma^3 + B_4 \gamma^2 + A_4 \gamma + 1 =0$ are given explicitly by the following expressions:
\be \label{4T-vep}
\vep (\boldsymbol{\theta}) = \pm \sqrt{\frac{4 - A_4 \pm \sqrt{A_4^2 - 4 B_4 + 8}}{8}},
\ee
where the $A_4$- and $B_4$-functions are defined by
\begin{align}
A_4 = &\, \mathsf{A}_0 + 2\sum_{j>0} \Re[\mathsf{A}_j e^{-i \theta_j}]+2\sum_{0<j< k}  \Re[\mathsf{A}_{jk} e^{-i \theta_{jk} }], \nonumber \\
B_4 = &\, \mathsf{B}_0 + 2\sum_{j>0} \Re[\mathsf{B}_j e^{-i \theta_j}]+2\sum_{0<j< k} \Re[\mathsf{B}_{jk} e^{-i \theta_{jk} }] \nonumber \\
      &\, + 2 \sum_{jkl \in P_{123}} \Re[\mathsf{B}_{jkl} e^{-i \theta_{jkl} } ]. 
\end{align}
Here we have used short-hand notations for phases $\theta_{jk} \equiv \theta_{j}-\theta_{k}$, and $\theta_{jkl} \equiv \theta_j+ \theta_k - \theta_l$, permutations $P_{123} \in \{123,312,231\}$, and $\Re[\cdot]$ denotes the real part of a complex number. Additionally, parameters $\mathsf{A}$ and $\mathsf{B}$ are functions of the scattering matrix elements $\{s_{jk}\}$. Specifically,
\begin{align}
\mathsf{A}_0 = \sum_{j=0}^{3} |s_{jj}|^2, \quad \mathsf{A}_j =s_{0j}^\ast s_{j0} , \quad \mathsf{A}_{jk} = s_{kj}^\ast s_{jk},
\end{align}
and
\begin{align}
& \mathsf{B}_0 =    \sum_{j<k} |\mathsf{s}_{jj;kk}|^2, \quad \mathsf{B}_j =    \sum_{k \neq 0,j} \mathsf{s}_{0k;kj}^\ast \mathsf{s}_{jk;k0},  \nonumber \\
& \mathsf{B}_{jk} = \sum_{l \neq j,k} \mathsf{s}_{kl;lj}^\ast \mathsf{s}_{jl;lk}, \quad 
  \mathsf{B}_{jkl} = \mathsf{s}_{0k;lj}^\ast \mathsf{s}_{k0;jl},
\end{align}
with $\mathsf{s}_{jk;lq} \equiv s_{jk}s_{lq}-s_{jq}s_{lk}$. The scattering matrix is parametrized by sixteen real parameters as in Ref.~\cite{Dita}, where $\{a,b,c,d,h,f\} \in [0,1]$ and $\varphi_{00,11,01,10,02,20,03,30,12,21} \in [0,2\pi]$. An inspection of these expressions reveals that despite the fact that we need ten independent phases to parametrize the scattering matrix, only six effective angles, $\phi_{1} \equiv \varphi_{12}-\varphi_{21}$, $\phi_{2} \equiv \varphi_{13}-\varphi_{31}$, $\phi_{3} \equiv \varphi_{14}-\varphi_{41}$, $\phi_{4} \equiv \varphi_{22}-\varphi_{23}$, $\phi_{5} \equiv \varphi_{22}-\varphi_{32}$, and $\phi_{6} \equiv \varphi_{11}+\varphi_{22}-\varphi_{12}-\varphi_{21}$, affect the Andreev spectrum in Eq.~\eqref{4T-vep}. The zero-energy states are determined by the $n=4$ Pfaffian equation \eqref{pf},
\be \label{pf4}
\mathrm{Pf}_4(\boldsymbol{\theta}) = \sum_{jkl \in P_{123}} 
\big[\mathsf{s}_{0k;lj}e^{i\theta_{jkl}/2}+\mathsf{s}_{jl;k0} e^{-i\theta_{jkl}/2}\big]= 0.
\ee
Via the unitary condition of $\hat{s}$, Eq.~\eqref{pf4} implies that $\sum_{jkl \in P_{123}} C_{jkl} \cos[(\theta_{jkl}-\zeta_{jkl})/2]=0$, where $C_{ijk}$ and $\zeta_{ijk}$ are real functions of $\{s_{jk}\}$.
Most importantly, this determines a Majorana-crossing surface in $\boldsymbol{\theta}$ space given by $\theta_{jkl}=\zeta_{jkl} \pm \frac{\pi}{2} (\mathrm{mod} \, 2 \pi)$. 

As a practical example, we study the energy bands of this model for the choice of incommensurate parameters: $a=1/4$, $b=1/\sqrt{3}$, $c=1/5$, $d=1/2$, $f=1/3$, $h=4/5$,  $\phi_{1} = \phi_2=\pi$, $\phi_3=0$, $\phi_4=\phi_5=-\pi/3$, and $\phi_6=\pi/6$. The energy spectrum, corresponding Chern number, and Josephson currents are shown in Fig.~\ref{fig-4T}. We observe that the lower bands exhibit $4\pi$-periodicity due to the Majorana crossings described by Eq.~\eqref{pf4}. Moreover, at $\theta_3 = \theta_3^\ast \approx 0.6990$ [Fig.~\ref{fig-4T}(b)] and $5.7883$ [Fig.~\ref{fig-4T}(c)] finite-energy Weyl nodes form at $(\theta_1^\ast,\theta_2^\ast) \approx (3.7785,3.2853)$ and $(3.2046,3.0051)$, respectively, between one of the higher and lower bands. The appearance of these nodal points is signaled by a change in the Chern number [Fig.~\ref{fig-4T}(a)]. At this point we comment that it was recently shown that such Majorana-Weyl crossings occur in a different model of a four-terminal junction formed between the end-states of one-dimensional topological superconductors (TS) of class D \cite{Meyer}. It has been pointed out that a finite Chern number $C_{ij}$ in this regime is associated with a quantized transconductance $G_{ij}=(2e^2/h)C_{ij}$. We confirm this result in our scattering matrix model and remark that the extra phase transitions in Fig.~\ref{fig-4T}(a) are related to gap closing/reopening at the band edge $\vep=1$ that may not be stable since the higher bands can strongly hybridize with the continuum $\vep>1$. 

The one-dimensional cut of the spectrum in Fig.~\ref{fig-4T}(b) along $\theta_1$ at $\theta_2=\theta_2^\ast\approx 3.2853$ is shown in Fig.~\ref{fig-4T}(d). The Josephson currents $J_{1,2}$ as functions of $\theta_{1,2}$ corresponding to the spectrum in Fig.~\ref{fig-4T}(b) are shown in Fig.~\ref{fig-4T}(e), where the hedgehog-like singularities are present at the Weyl nodes $(\theta_1^\ast + 2 \pi,\theta_2^\ast)$ and $(\theta_1^\ast,\theta_2^\ast+2 \pi)$. We note that the other two nodal points at $(\theta_1^\ast,\theta_2^\ast)$ and $(\theta_1^\ast+2 \pi,\theta_2^\ast+2 \pi)$ do not induce current singularities since the higher- and lower-band contributions cancel each other. This can be observed in Fig.~\ref{fig-4T}(f), which displays $J_1(\theta_1)$ along the cut at $\theta_2 = \theta_2^\ast$. 

\begin{figure}
\includegraphics[width=0.24\textwidth]{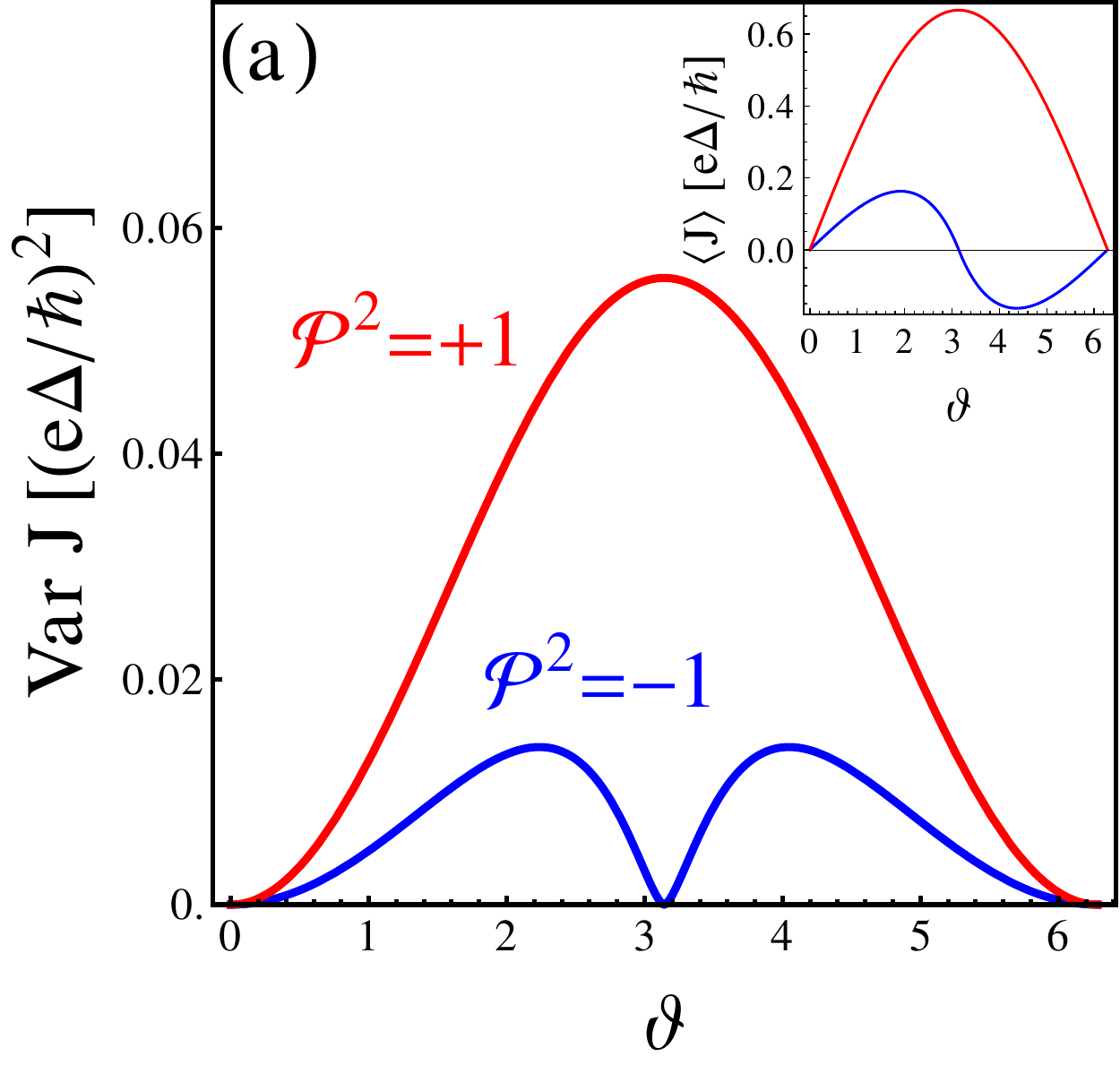}
\includegraphics[width=0.23\textwidth]{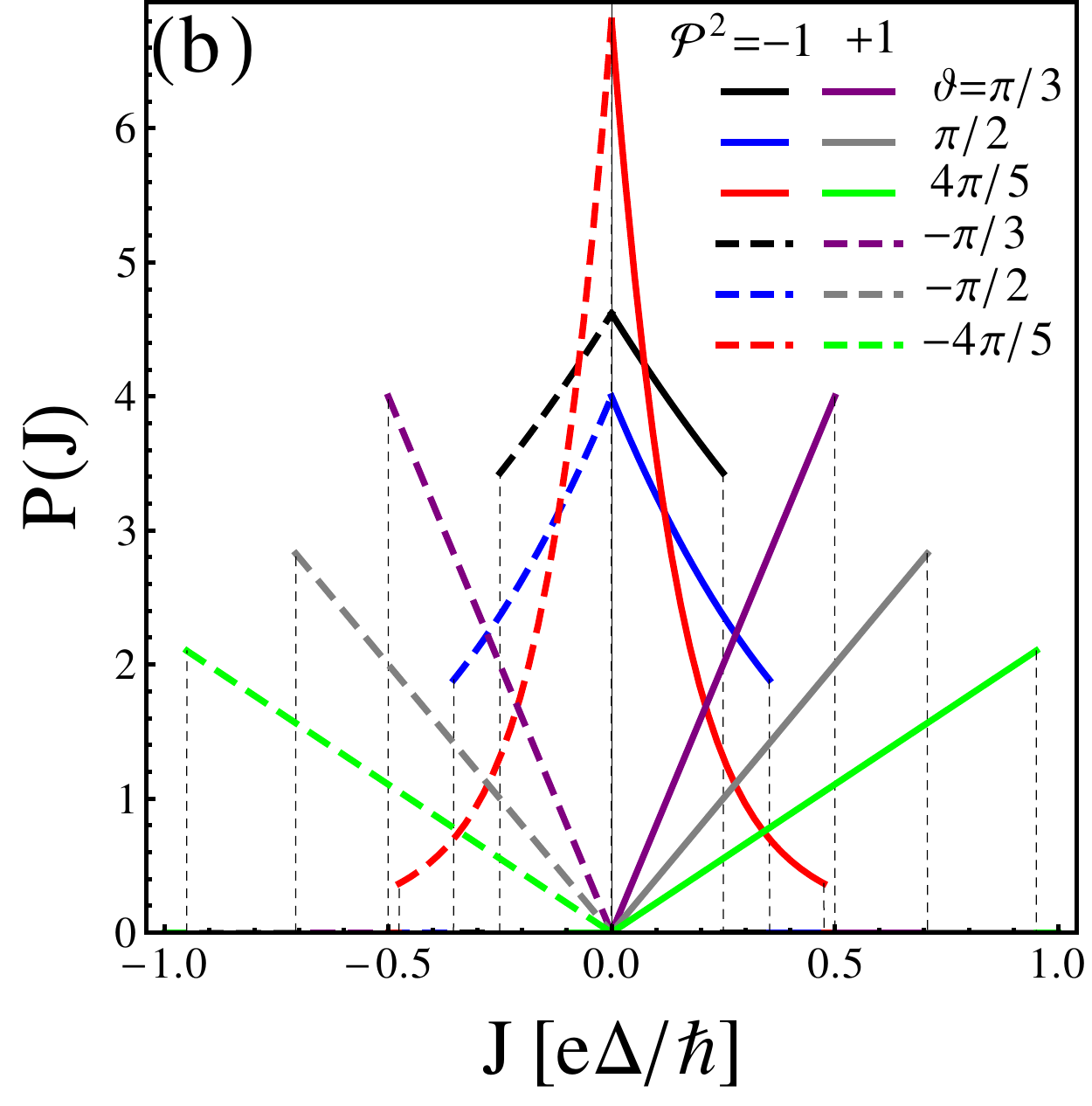}
\caption{[Color online] Josephson-current statistics of the two-terminal junctions. (a) Josephson-current variance $\mathrm{Var}J$ as a function of phase variable $\vartheta$ [Eq.~\eqref{2TVarJ}]. The inserted panel shows the expectation value $\langle J\rangle$. (b) Probability distribution function $P(J)$ for various phase variables $\vartheta$ [Eq.~\eqref{2TPJ}]. } 
\label{fig-2Tpj}
\end{figure}

\section{Fluctuations in topological junctions}\label{sec:fluctuations}

In the previous section we studied Josephson current for the given realization of the scattering matrix. As alluded to in the Introduction, this current is expected to display reproducible sample-to-sample fluctuations and it is thus of interest to study its statistical properties. We primarily focus on its variance and also on the cross-correlation function that can be experimentally accessed in the multiterminal devices. As is known from quantum transport theoretical approaches,     statistical transport properties of phase-coherent mesoscopic systems can be conveniently computed by means of averaging over a random-matrix that describes the system. In open systems, the averaging is done over the scattering matrix and one typically considers two models of junctions: chaotic cavities or disordered contacts. The former case is more suitable for the model considered in this work. We thus follow classical works by Baranger and Mello \cite{Baranger}, and Jalabert, Pichard, and Beenakker \cite{Jalabert} who studied conduction through a chaotic cavity on the assumption that the scattering matrix is uniformly distributed in the unitary group, restricted only by symmetry. This is the circular ensemble, introduced by Dyson, and shown to apply to a chaotic cavity by Blumel and Smilansky \cite{Smilansky}. In other words we consider a SNS junction where the normal region is a chaotic quantum dot \cite{Whisler}.  

\begin{figure}
\includegraphics[width=0.23\textwidth]{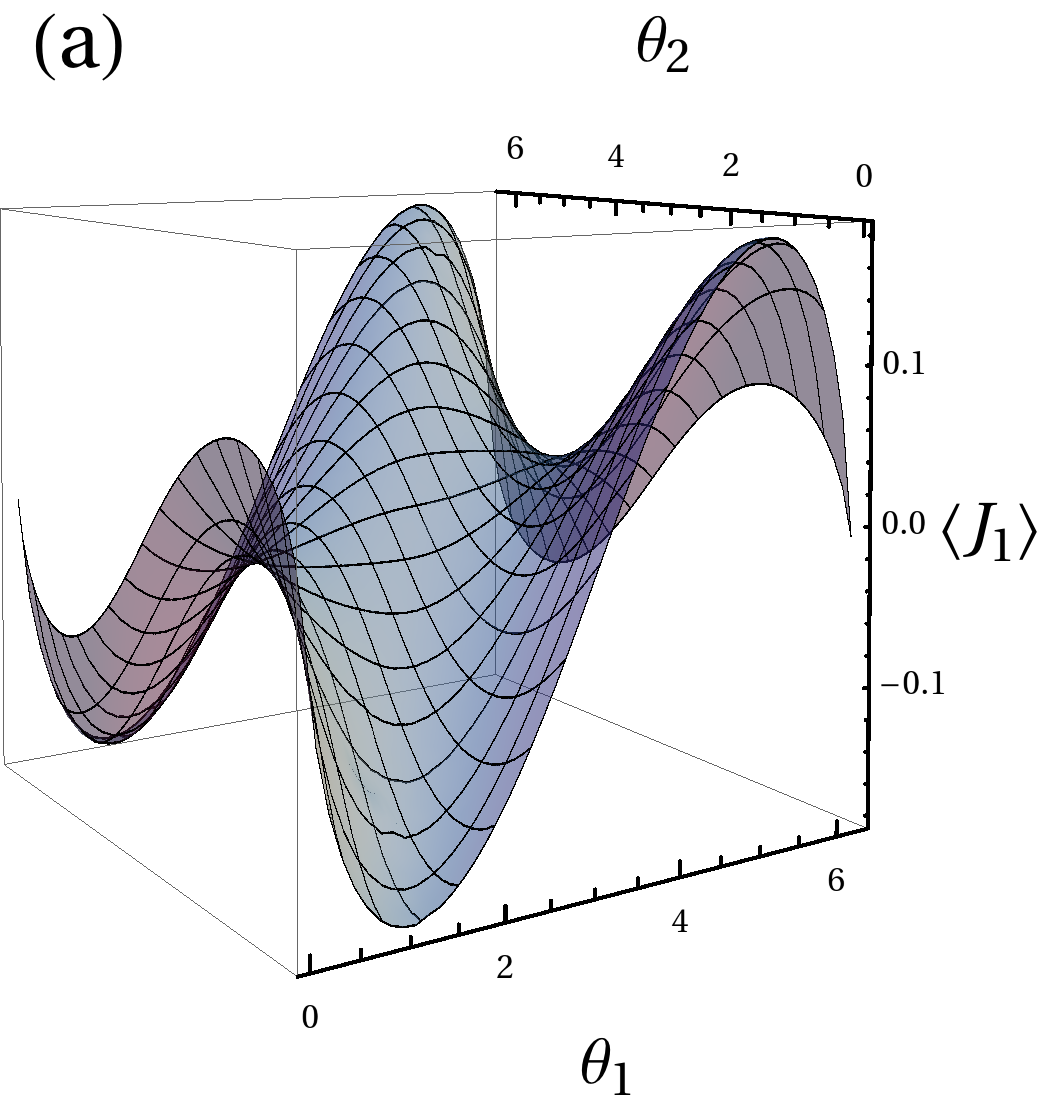} \,
\includegraphics[width=0.23\textwidth]{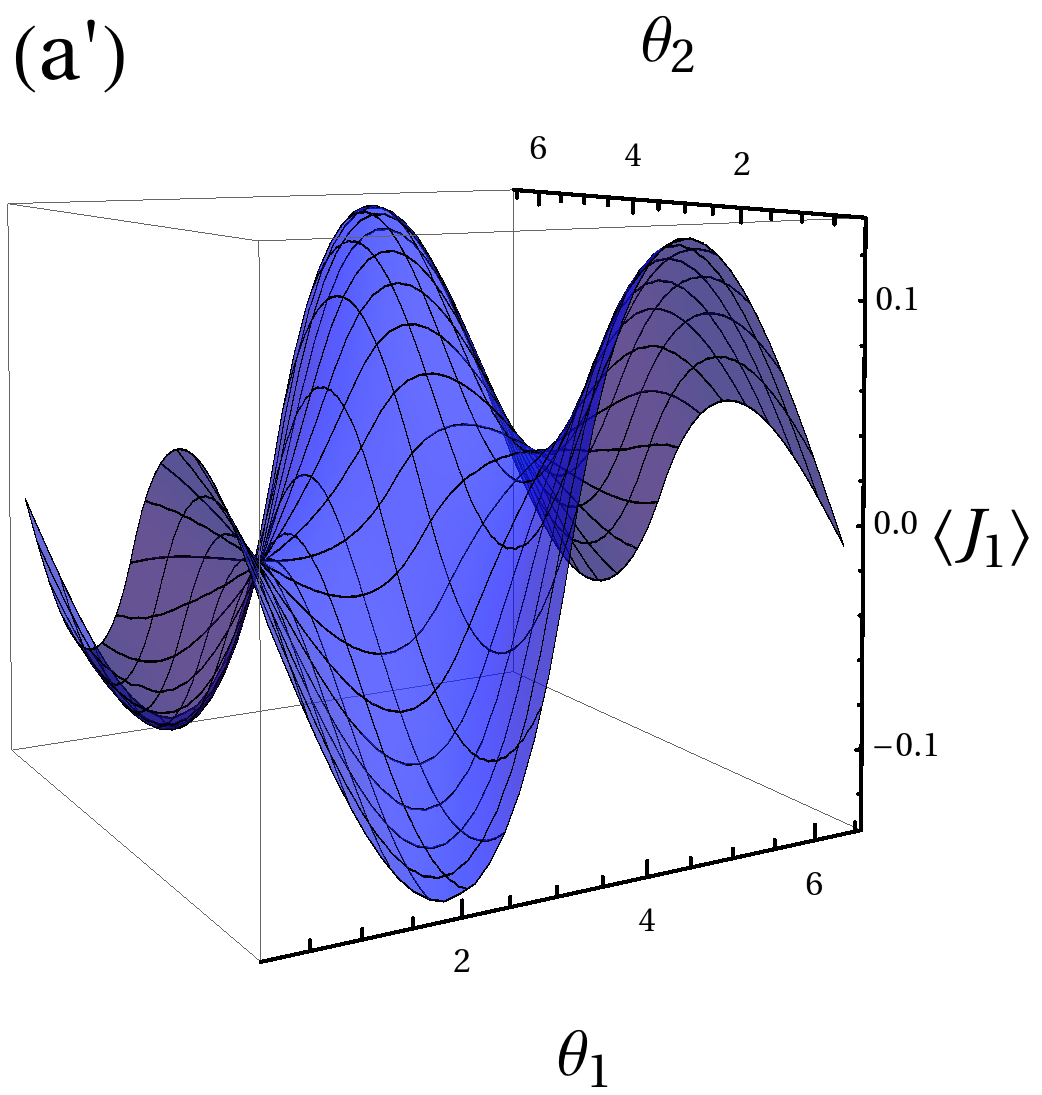} \\
\includegraphics[width=0.23\textwidth]{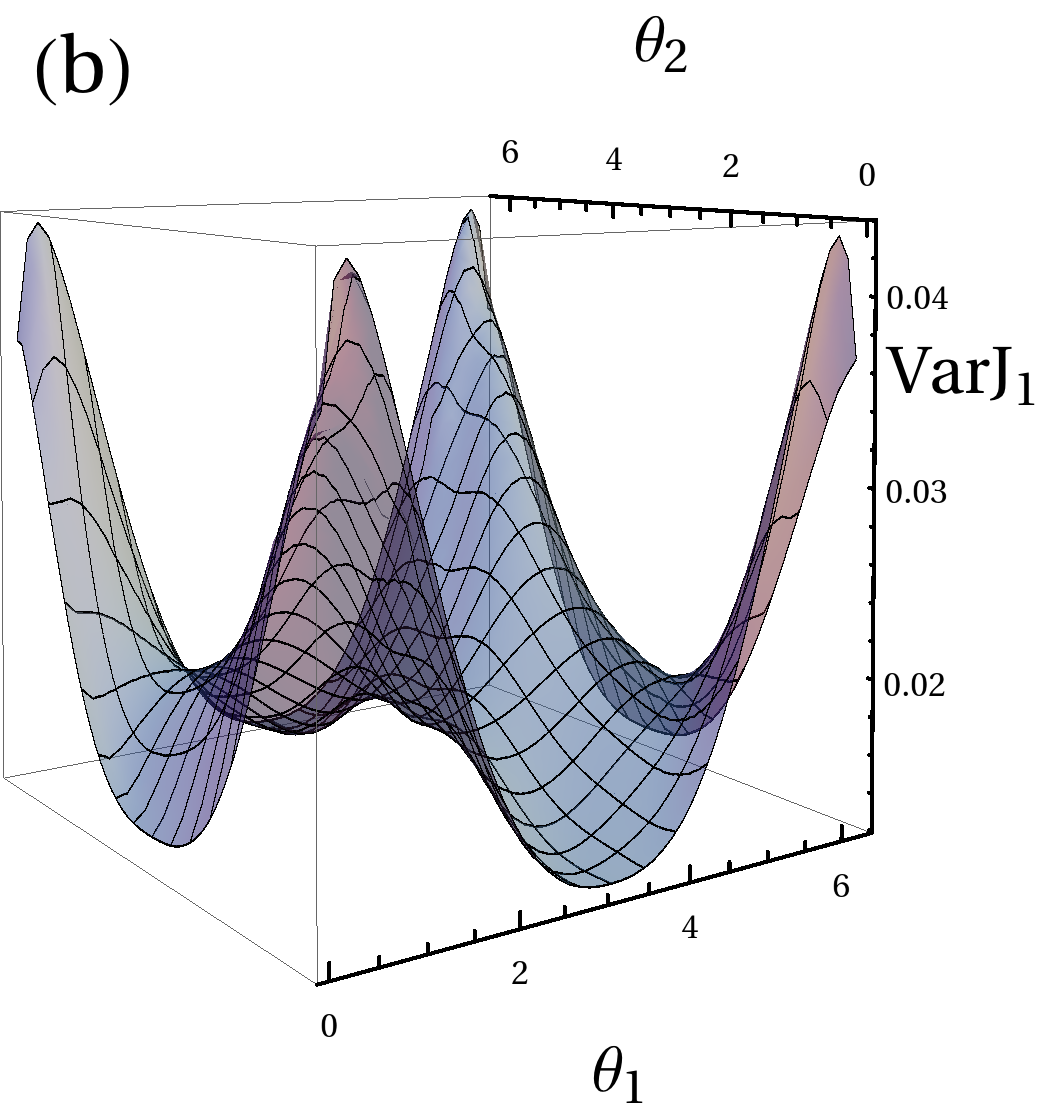} \,
\includegraphics[width=0.23\textwidth]{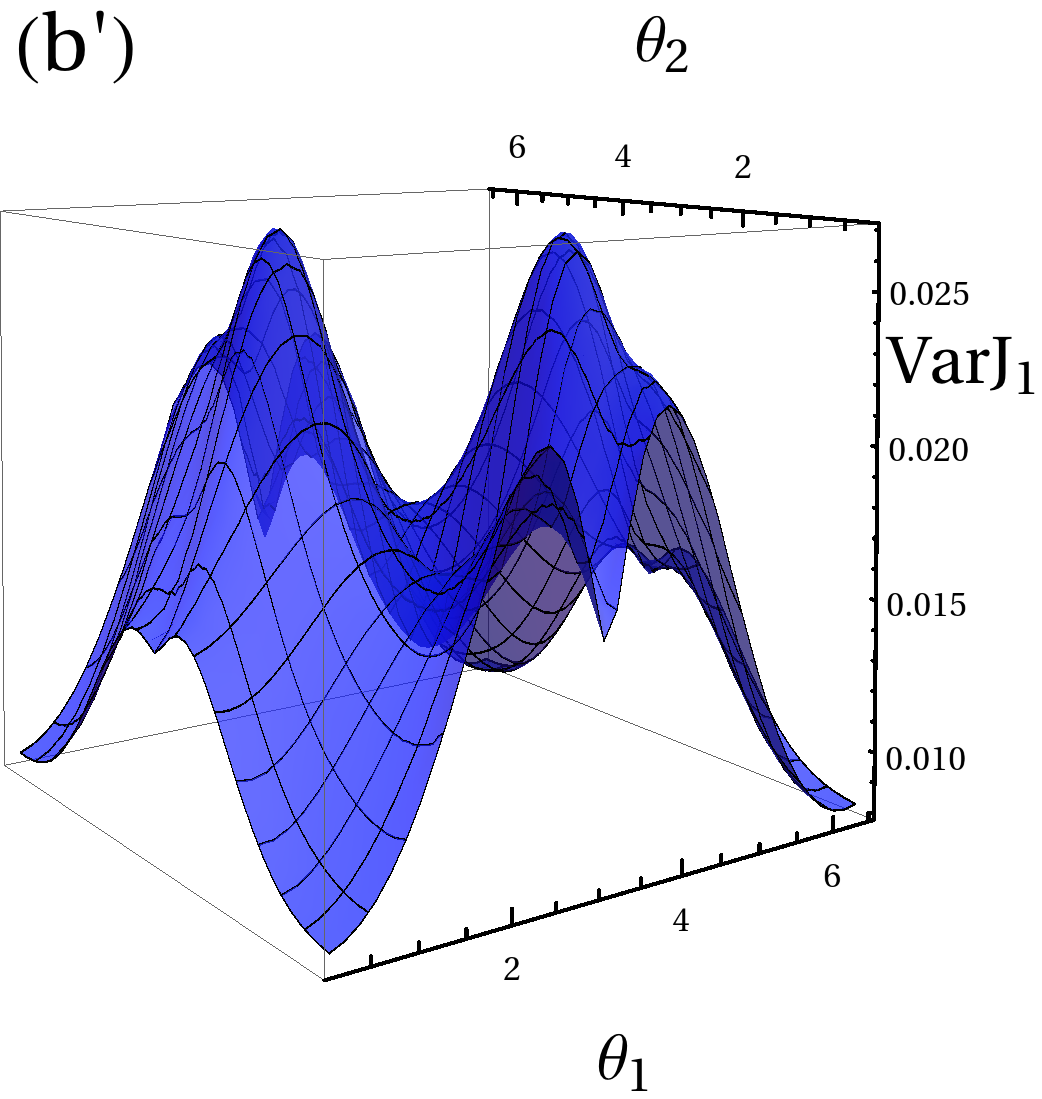} \\
\includegraphics[width=0.23\textwidth]{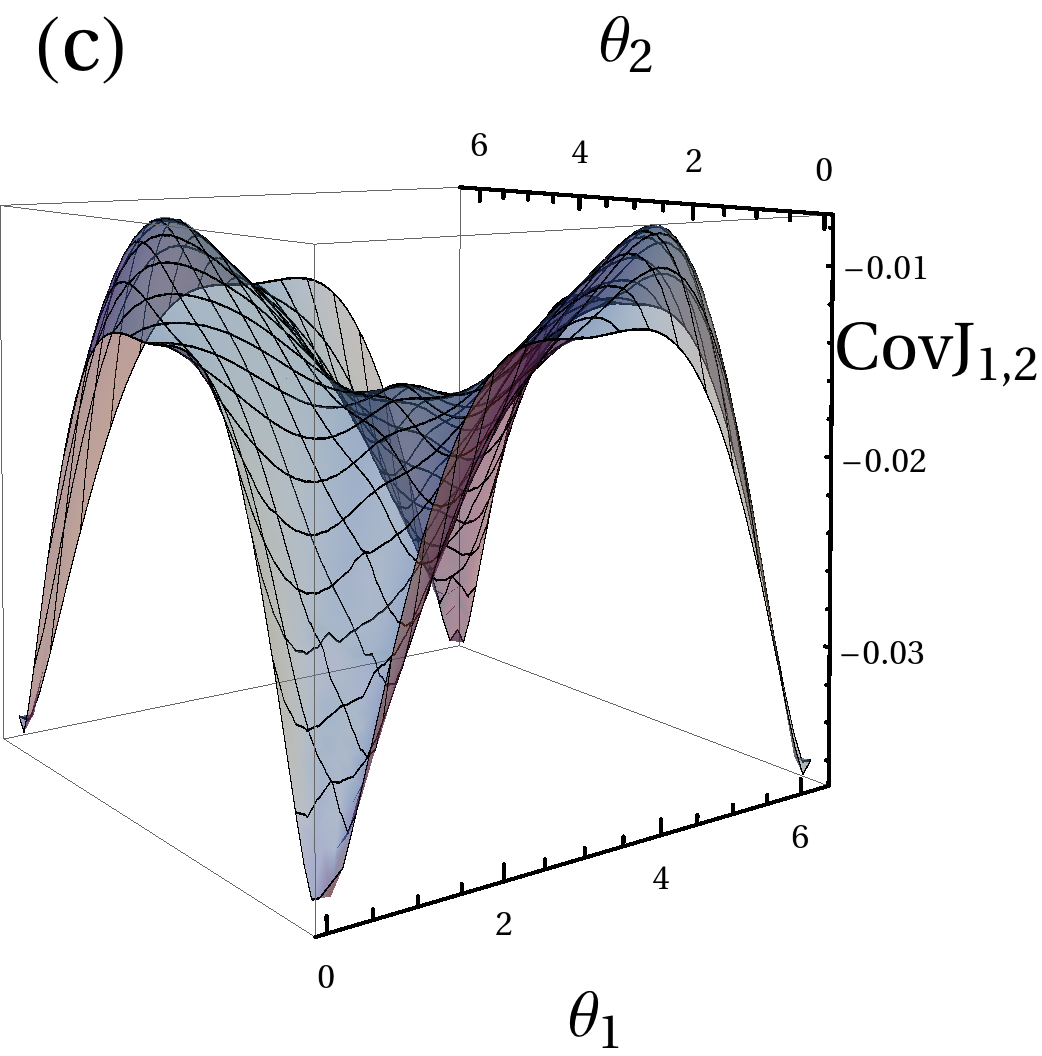} \,
\includegraphics[width=0.23\textwidth]{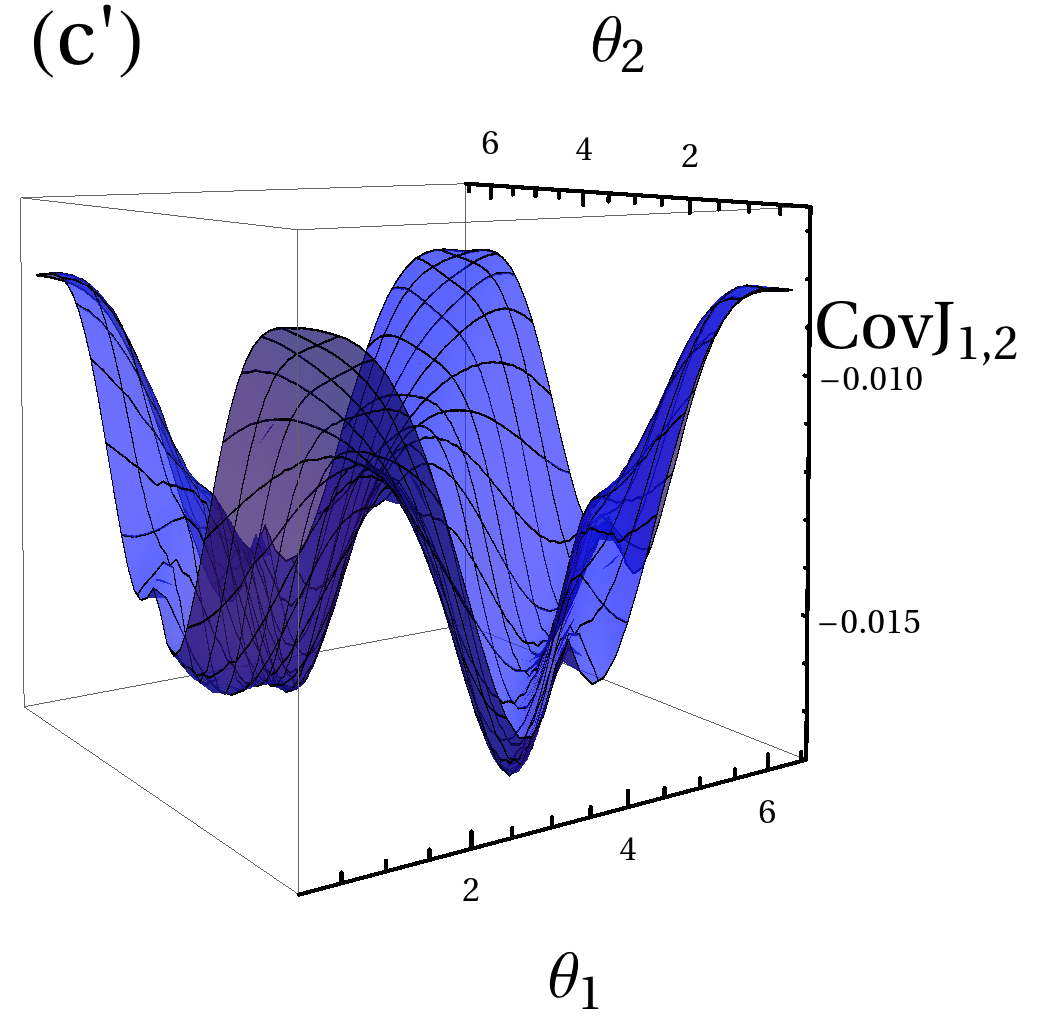}
\caption{[Color online] Josephson current statistics of three-terminal junctions. Panels (a, b, c) and (a', b', c') display the results for $\mathcal{P}^2=+1$ and $\mathcal{P}^2=-1$ junctions, respectively. (a) and (a') Expectation of $J_1$ as a function of $\vartheta_{1,2}$. (b) and (b') Variance of $J_1$ as a function of $\vartheta_{1,2}$. (c) and (c') Covariance of $J_{1,2}$ as a function of of $\vartheta_{1,2}$.} 
\label{fig-3T-stat}
\end{figure}

The probability density $\rho(\vec{x})$, an invariant Haar measure, of the $\hat{s}$-matrix parameters $\vec{x}$ is given by 
\be \label{def-rho}
\rho(\vec{x}) \equiv \sqrt{|\mathrm{Det} \hat{M}(\vec{x})|}, \quad M_{\mu \nu} \equiv \sum_{i,j} \frac{\partial s_{ij}}{\partial x_\mu} \frac{\partial s_{ij}^\ast}{\partial x_\nu}.
\ee
The distribution function of an observable $Q(\vec{x})$, defined as $P(Q) \equiv \int \, d \vec{x} \rho(\vec{x}) \delta(Q-Q(\vec{x}))$, is in practice calculated by the characteristic function 
\be  \label{def-Pp}
p(\lambda) = \langle e^{i\lambda Q(\vec{x})} \rangle, \quad P(Q) = \frac{1}{2 \pi}\int\! d\lambda e^{-i \lambda Q} p(\lambda),
\ee
where $\langle \cdots \rangle \equiv \int\!d\vec{x} \rho(\vec{x}) (\cdots)$, denoting the circular unitary ensemble (CUE) average.

For a benchmark we first study the statistics of the Josephson current $J(\vartheta)$ in the two-terminal junctions and take $2 \Delta/\hbar$ as the units of $J$ in the following discussion. From Eqs.~\eqref{s2} and ~\eqref{def-rho} we obtain a constant invariant measure $\rho(T)=1$. We recall that this simplicity is specific to the unitary case; for instance, in orthogonal symmetry the probability density is not flat in $T$ even for a single-channel limit. All the moments as well as the distribution function of the Josephson current can be obtained analytically. The $m$-moment is given by the expression
\begin{subequations} 
\begin{align}
\langle J^{m} \rangle = &\, \frac{2}{m+2}\sin^{m}\left( \frac{\vartheta}{2} \right), \quad \mathcal{P}^2 = +1, \\
\langle J^{m} \rangle = &\, \frac{1}{m+1}\left(\frac{\sin{\vartheta}}{4}\right)^m \nonumber \\ 
 & \times {}_2F_1 \left[ \frac{m}{2}, m+1; m+2; \sin^2\left(\frac{\vartheta}{2}\right) \right], \quad \mathcal{P}^2 = -1,
\end{align}
\end{subequations}
where $m \in \mathbb{N}$ and ${}_2F_1 (a, b; c; z)$ is the hypergeometric function. Therefore, the variance for $ \mathcal{P}^2 = +1$ reads 
\begin{subequations} \label{2TVarJ}
\begin{align}
\mathrm{Var}\, J = &\, \frac{1}{18} \sin^{2}\left( \frac{\vartheta}{2} \right),
\end{align}
whereas for the non-topological case $ \mathcal{P}^2 = -1$ it has a different look, 
\begin{align}
\mathrm{Var}\, J = &\, \frac{\sin^2{\vartheta}}{192} \left[ 1 + \sin^2\left(\frac{\vartheta}{2}\right) + \frac{113}{120}\sin^4\left(\frac{\vartheta}{2}\right) + \cdots \right],
\end{align}
\end{subequations}
as depicted in Fig.~\ref{fig-2Tpj}(a). In the topological regime the variance inherits $4\pi$ periodicity and has a remarkably simple form. In the non-topological regime, our result is similar to that of Chalker and Mac\^edo \cite{Chalker}, albeit with different numerical coefficients as they considered the multi-mode disordered junction model where averaging is done over the Dorokhov distribution of transmission eigenvalues. Finally, the Josephson-current distribution function takes the form, for $\mathcal{P}^2 = +1$,
\begin{subequations} \label{2TPJ}
\begin{align}
P(J;\vartheta) =  \frac{2 J}{J_\mathrm{c}^+|J_\mathrm{c}^+|} 
\Theta \left( |J_\mathrm{c}^{+}|-\left|2 J - J_\mathrm{c}^{+} \right|  \right), 
\end{align}
and, for $\mathcal{P}^2 = -1$,
\begin{align}
P(J;\vartheta) =  \frac{8 \, \Theta \left( |J_\mathrm{c}^{-}|-\left|2 J - J_\mathrm{c}^{-} \right|  \right)}{\left|\sin{\vartheta} \right| K \left( J \tan\frac{\vartheta}{2}  \right)  \left[1+K^2\left(  J \tan\frac{\vartheta}{2} \right)\right]}, 
\end{align} 
\end{subequations}
where $J_\mathrm{c}^+(\vartheta) = \sin(\vartheta/2)$ and $J_\mathrm{c}^{-}(\vartheta)= \sin(\vartheta/2) \, \mathrm{sgn}[\cos(\vartheta/2)]/2$ are the critical currents, $\Theta(x)$ is the Heaviside step function, and the function $K(x) =|x|+\sqrt{1+x^2}$. As shown in Fig.~\ref{fig-2Tpj}(b), in both $\mathcal{P}^2=\pm 1$ classes the relation $P(J;-\vartheta)=P(-J;\vartheta)$ is satisfied. (i) For $\mathcal{P}^2=+1$, $P(J)$ is a linear function of $J$ for which $P(0)=0$ and the slope is defined by the phase $\vartheta$. In particular $P(J)= \delta(J)$ for $\vartheta = 2k\pi$ with $k \in \mathbb{Z}$. (ii) For $\mathcal{P}^2=-1$, $P(J)$ is smaller for a larger current amplitude and $P(J)= \delta(J)$ for $\vartheta = k\pi$ with $k \in \mathbb{Z}$.

We proceed to study the Josephson-current statistics of the three-terminal junctions. From Eq.~\eqref{def-rho} we obtain the probability density of the effective parameters $\vec{x} = (a,b,c,\varphi)$,
\be 
\rho(\vec{x}) = \mathcal{N} a b \sqrt{\frac{(2-a^2)[(1-a^2)(2-b^2)+a^2 b^2 \sin^2{\varphi}] }{(1-a^2)(1-b^2)(1-c^2)}},
\ee  
where $ \mathcal{N} \approx 1.7671 \times 10^{-2}$ is the normalization constant. The numerical results of the expectation value, variance, and covariance of $J_{1,2}(\vartheta_{1,2})$ are shown in Fig.~\ref{fig-3T-stat}, where the covariance is defined as $\mathrm{Cov} J_{1,2} \equiv \langle J_1 J_2 \rangle - \langle J_1 \rangle \langle J_2 \rangle$. The general relations $\langle J_1^m(\vartheta_{1},\vartheta_{2}) \rangle = \langle J_2^m(\vartheta_{2},\vartheta_{1})\rangle$ and $\mathrm{Cov}J_{1,2}(\vartheta_{1},\vartheta_{2})= \mathrm{Cov}J_{1,2} (\vartheta_{2},\vartheta_{1})$ are satisfied. We observe that the variances and covariances distinguish the $\mathcal{P}^2 = \pm 1$ junctions. 

\begin{figure}
\includegraphics[width=0.23\textwidth]{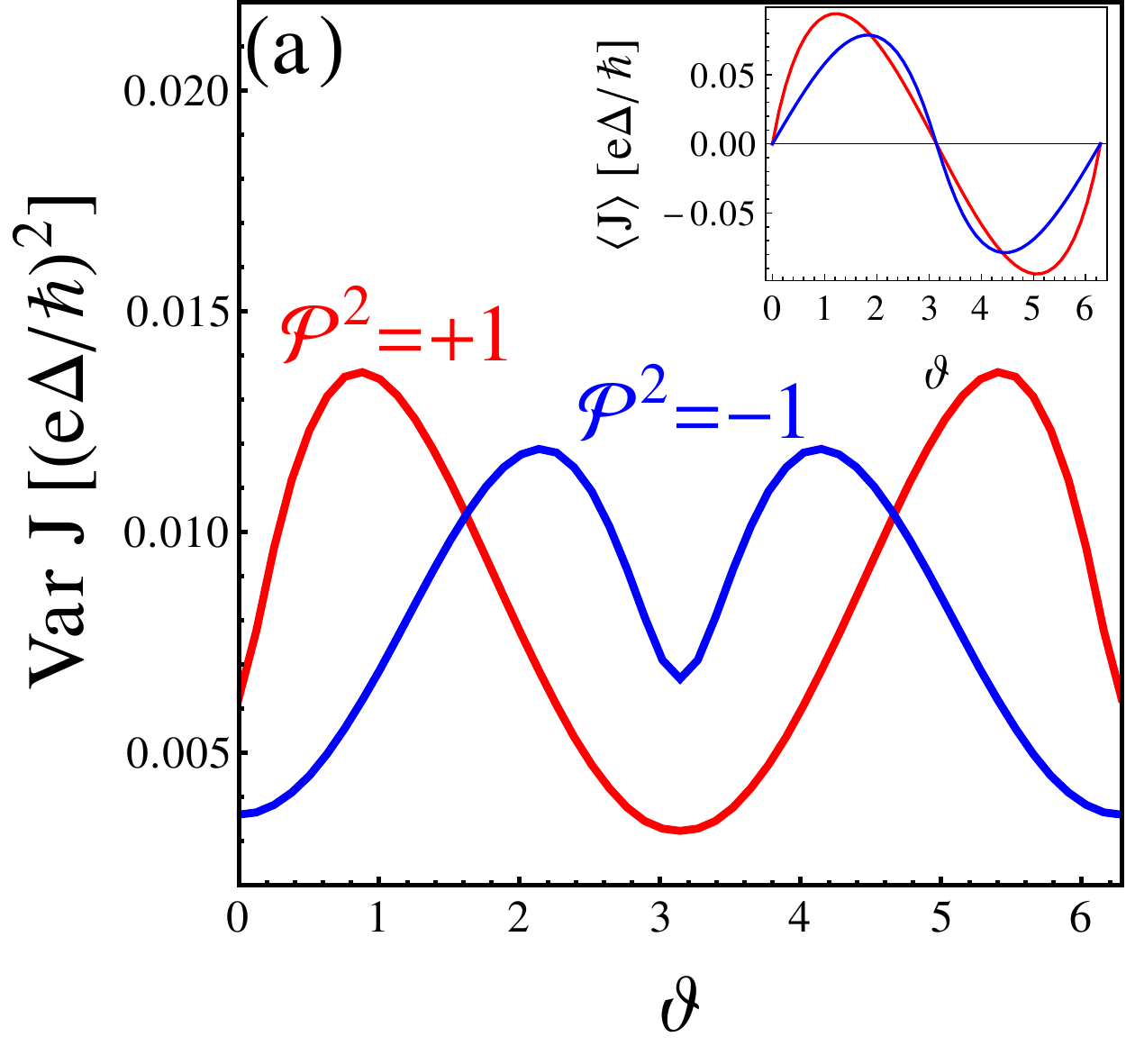}
\includegraphics[width=0.235\textwidth]{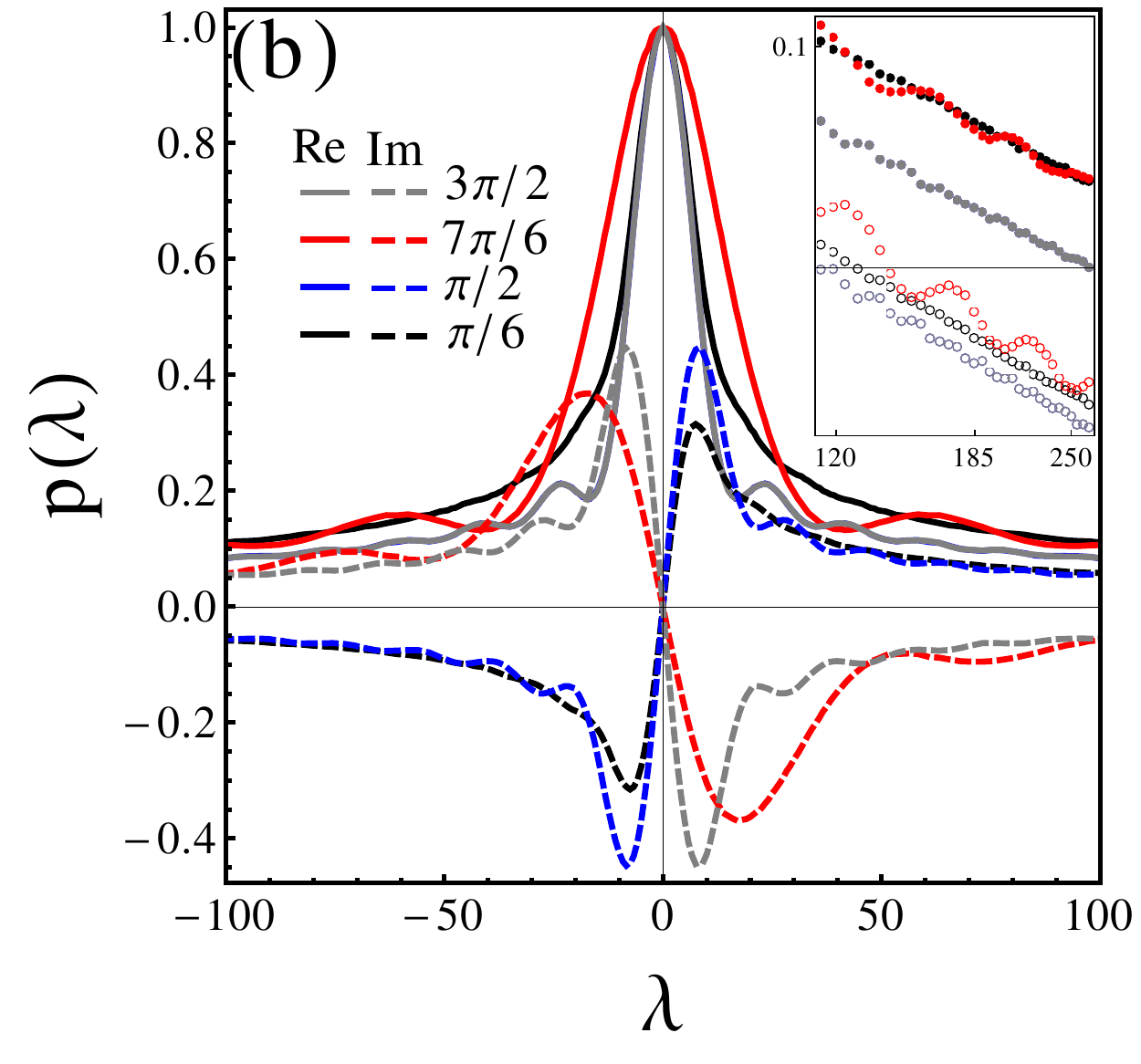} 
\caption{[Color online] Josephson current statistics of S-TS junctions. (a) Variance of $J$ as a function of $\vartheta$. We present the $\mathcal{P}^2=-1$ results for comparison. The insert panel shows the expectation value as a function of $\vartheta$. (b) Characteristic function $p(\lambda)$ for various values of $\vartheta$. The solid (dashed) lines are the real (imaginary) part of $p(\lambda)$. The insert panel is the log-log plot for $\lambda\gg 1$.} 
\label{fig-TST-stat}
\end{figure}

For $\mathcal{P}^2=+1$ three-terminal junctions, by integrating one and two terminals we can construct an effective two-terminal S-TS junction \cite{Ioselevich} which supports a single channel on one lead (topological), with phase $\theta_0=0$, and two channels on the other (conventional), with phase $\theta=\theta_1=\theta_2$. Defining $\vartheta = \vartheta_1$ and $\phi= \phi_2-\phi_1$ in Eqs.~\eqref{3T-spec} and \eqref{B-trsb}, we obtain the Josephson current through the two leads,
\be 
J(\vec{x}^\prime;\vartheta) = \frac{e \Delta}{\hbar} \frac{\partial_\theta B_3(\vartheta,\vartheta)}{4 \vep_{\pm}(\vartheta,\vartheta)},
\ee
which depends on five independent parameters $\vec{x}' = (a,b,c,\varphi,\phi)$. We present the numerical results of the statistical properties of $J(\vartheta)$ for such a configuration in Fig.~\ref{fig-TST-stat}. The expectation and variance of $J$ as functions of $\vartheta$ are shown in Fig.~\ref{fig-TST-stat}(a). The characteristic function $p(\lambda;\vartheta)$ for various $\vartheta$ is shown in Fig.~\ref{fig-TST-stat}(b). We observe that, for $\lambda \gg 1$, $p(\lambda) \propto \lambda^{-\alpha}$ with the exponent $\alpha$ being $\vartheta$-independent. We calculate the $\vartheta$-averaged characteristic function $\bar{p}(\lambda) \equiv \frac{1}{2 \pi}\int_{0}^{2 \pi} \! d\vartheta p(\lambda;\vartheta)$ and fix the exponent $\alpha \sim 0.55$. In particular, this analysis enables us to extract the asymptotic behavior of the full distribution function $P(J)$ which is found to exhibit a universal power-law scaling $P(J) \propto J^{-(1-\alpha)}$ in the limit $J\ll1$. 
 
\section{Discussion and outlook}

In this work we applied methods of scattering matrix theory to study the transport properties of multiterminal Josephson junctions of topological superconductors. We have examined the spectrum of sub-gap states in two-, three-, and four-terminal configurations and determined that the texture of resulting Andreev bands in the multidimensional parameter space of superconducting phases can produce nonvanishing fluxes of the Berry curvature. These properties translate into the quantized nonlocal conductances of these devices. We have also studied the current-phase relationships and interaction of supercurrents, as well as their mesoscopic statistical properties. In particular, we discussed the universal regime of current fluctuations and computed supercurrent variance as well as the current cross-correlation function in the topological regime. We close this work with a few comments in relation to existing and possible future experiments in which the fundamental physics of multiterminal Josephsonic devices could be further explored. 

Recently, Josephson supercurrent and conductance were measured as a function of geometry, temperature, and gate voltage in proximitized planar junction devices composed of superconductors and surface states of a topological insulator (S-TI-S junctions) in order to determine the nature of the electronic transport in these systems. The supercurrent was found to exhibit a sharp drop as a function of gate voltage, see Figs. (2) and (6) of Ref. \cite{Kurter-PRB}, superimposed with reproducible noise whose magnitude was a fraction of the critical current. The systematic trend in the critical current dependence was explained by a mechanism related to the relocation of the topological surface state with respect to trivial conducting two-dimensional states formed by band-banding near the surface. In real space, a negative gating potential pushes the trivial state below the topological surface states, exposing the topological state to the disordered surface of the TI. As a result, the magnitude of the supercurrent changes sharply. The noise was attributed to the percolation effects as near the voltage threshold it is likely that local charge fluctuations cause the path of the supercurrent to be highly meandering. We wish to point out that there is possibly an alternative picture as this noise could be of mesoscopic origin. This evidence is further supported by observed similar reproducible noise features in Fraunhofer magneto-oscillations of the critical current. While our model is not directly applicable to S-TI-S junctions we make the observation that the magnitude of current fluctuations is consistent with the expectations that disorder scattering causes observed mesoscopic effects.   

In addition, we wish to nore that related statistical properties of supercurrents can also be studied by measuring switching current distributions. In particular, for topological Josephson devices, the critical current measurements can potentially enable determining the parity state of a Majorana fermion (pair) in a junction since the supercurrent acquires an anomalous fractional component due to Majorana modes, $\pm \sin(\vartheta/2)$, where the sign encodes the parity. The typical switching measurement is performed by ramping the bias current through the junction to detect the current value at which the junction jumps to the finite voltage state. By repeating this protocol many times and accumulating statistics of random supercurrent switching events (as previously successfully implemented in various mesoscopic proximity circuits, e.g. nanowires and graphene layers \cite{Coskun,Murphy}), one expects to reveal a bimodal distribution indicating the two parity states. If the separation of the two peaks in this distribution is wide enough, one can detect (with some fidelity) the parity state. Mutiterminal devices considered in this work can provide an actual hardware platform to conduct such experiments and our transport theory will be useful in modeling future measurements. In particular, knowledge of the current-phase relationship is needed for determining the energy barrier of a phase slip that triggers the switching. Furthermore, these developments are also inspired by the potential application of multiterminal devices in the design of protected superconducting qubits.

\section*{Acknowledgment}

We would like to thank Dale Van Harlingen for the discussions, especially on the topic of the gate dependence of the critical current fluctuations in topological Josephson junctions \cite{Kurter-PRB} and the feasibility of switching current experiments in the tri-junction configuration. The work of H.-Y. X was supported by National Science Foundation Grants No. DMR-1653661 and DMR-1743986. The work of A.L. was supported in part by the U.S. Department of Energy, Office of Science, Basic Energy Sciences, under Award No. DESC0017888, and the Army Research Office, Laboratories for Physical Sciences Grant No. W911NF-18-1-0115. H.-Y. X. acknowledges the hospitality of the Kavli Institute of Theoretical Sciences (KITS) where this work was partially done.

\end{document}